\documentclass[12pt]{article}
\usepackage[T1]{fontenc}
\usepackage[utf8]{inputenc}
\usepackage[english]{babel}
\usepackage{amsmath} 
\usepackage{graphicx} 
\usepackage{caption} 
\usepackage{subfig} 
\usepackage{amssymb}
\usepackage[dvipsnames]{xcolor} 
\usepackage{float}
\usepackage{slashed} 
\usepackage{bm} 
\usepackage{hyperref} 
\usepackage{comment} 
\usepackage{cite} 
\usepackage{multirow}
\usepackage{bbold}
\usepackage[normalem]{ulem}  

\newcommand{\LamQCD}{\Lambda_{\rm QCD}}
\newcommand{\as}{\alpha_s}

\newcommand{\red}[1]{{\color{red}#1}}
\newcommand{\LambdaBNV}{\Lambda_{\rm BNV}}

\newcounter{MBQ}

\begin{document}
\allowdisplaybreaks

\begin{titlepage}

\begin{flushright}
{\small
TUM-HEP-1504/24\\
USC-TH-2024-01\\
April 10, 2024 \\
}
\end{flushright}

\vskip1cm
\begin{center}
{\Large \bf\boldmath Indirect constraints on third generation \\[0.1cm]baryon number violation}
\end{center}

\vspace{0.5cm}
\begin{center}
{\sc Martin~Beneke,$^a$ 
Gael Finauri,$^{a}$ Alexey A. Petrov$^{b}$} \\[6mm]
{\it $^a$Physik Department T31,\\
James-Franck-Stra\ss{}e~1, 
Technische Universit\"at M\"unchen,\\
D--85748 Garching, Germany}\\[0.3cm]

{\it $^b$Department of Physics and Astronomy, \\
University of South Carolina, Columbia, \\
South Carolina 29208, USA}\\[0.3cm]
\end{center}

\vspace{0.6cm}
\begin{abstract}
\vskip0.2cm\noindent
The non-observation of baryon number violation suggests 
that the scale of baryon-number violating interactions at zero temperature is comparable to the GUT scale. However, the pertinent measurements involve hadrons made of the first-generation quarks, such as protons and neutrons.
One may therefore entertain the idea that new 
flavour physics breaks baryon number at a much lower scale, but only in the coupling to a third generation quark, leading to observable baryon-number violating $b$-hadron decay rates. 
In this paper we show that indirect constraints on the new physics scale $\LambdaBNV$ from the existing bounds on the proton lifetime do not allow for this possibility. For this purpose we consider the three dominant proton decay channels $p \to \ell^+ \nu_\ell \bar{\nu}$, $p \to \pi^+ \bar{\nu}$ and $p \to \pi^0 \ell^+$ mediated by a virtual bottom quark.
\end{abstract}
\end{titlepage}

\tableofcontents
\newpage

\section{Introduction}

The presence of baryon-number violating processes is an essential part of the generation of matter-antimatter asymmetry of the Universe. 
While a baryon-number violating mechanism is realized in the Standard Model (SM), it only appreciably affects processes in the early Universe. Strong constraints on baryon number violation (BNV) today arise from the lifetime $\tau_p$ of the proton, which is constrained experimentally to $\tau_p \ge 10^{30}-10^{34}~\text{yr}$~\cite{PDG}. This is usually interpreted as an indication that the accidental baryon number symmetry of the SM can be explicitly broken only at a very high scale of $\mathcal{O}(10^{15-16}~\text{GeV})$.

New physics (NP) scenarios often contain additional ways to violate 
baryon number conservation. It is not excluded 
that the breaking of baryon number symmetry is quark- and 
lepton-flavour dependent. Discrepancies between measurements 
and theoretical predictions have been observed in the heavy-flavour sector in recent years~\cite{Albrecht:2021tul,Capdevila:2023yhq}. If not explained by QCD effects or experimental systematics, the 
required new flavour physics usually invokes highly specific generation-dependent couplings at the scale $1.5 - 5$~TeV, 
which in principle could be baryon-number violating. 
In fact, explicit models have been built \cite{Baldes:2011mh,Nelson:2019fln,Elor:2018twp} in which BNV could occur at a similar scale, but only when 
the third quark family is involved. One may then speculate that BNV might show up in $b$-hadron decays with detectable rates. 
In this paper, we note that even if BNV is confined to operators containing third-generation quarks, 
the proton can still decay through virtual heavy quarks by a combination 
of electroweak charged (or neutral) current and BNV interactions. 
This leads to lower bounds 
on the third-generation BNV scale $\LambdaBNV$, 
which exclude detectable BNV $b$-decays by a large margin.

\begin{figure}
\centering
\includegraphics[width=1\textwidth]{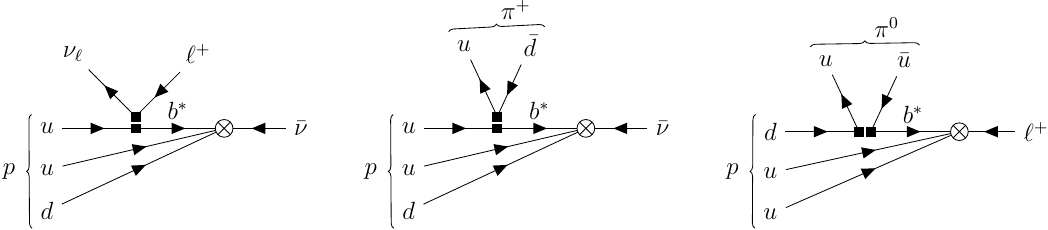}
\caption{\small Examples of tree-level partonic diagrams for the decays $p \to \ell^+ \nu_\ell \bar{\nu}$, $p \to \pi^+ \bar{\nu}$ and $p \to \pi^0 \ell^+$. The square dots represent the weak effective vertices while $\otimes$ stands for the four-fermion BNV SMEFT operator insertion.}
\label{fig:diags1}
\end{figure}

We consider the three simplest and dominant proton decay channels: first the purely leptonic $p\to \ell^+ \nu_\ell \bar{\nu}$ mode mediated by the weak transition  $u \to b^* \ell^+ \nu_\ell$. 
Second the two-body semi-leptonic channels $p \to \pi^+ \bar{\nu}$ and $p \to \pi^0 \ell^+$ mediated by $u \to b^* u \bar{d}$.  
Example tree-level diagrams are shown in Figure~\ref{fig:diags1}. The weak currents are combined with an insertion of the BNV SMEFT four-fermion operators containing a $b$ quark.\footnote{
A related idea was considered in~\cite{Marciano:1994bg,Hou:2005iu} with operators involving a $\tau$ lepton, as well as operators built \emph{only} from third generation fermions~\cite{Hou:2005iu}.} 
Our strategy is to make use of the hierarchy\footnote{$m_W$, $m_b$, $m_p$ denote the masses of the $W$-boson, bottom quark and proton, respectively.} $m_W \gg m_b \gg m_p \sim \LamQCD$ to integrate out possible virtual weak bosons and the virtual $b$-quark in order to obtain four- and six-fermion effective contact interactions.
From the theoretical calculation of the decay rates we then constrain the NP scale $\LambdaBNV$ from experimental bounds~\cite{Super-Kamiokande:2013rwg,Super-Kamiokande:2014pqx,Super-Kamiokande:2020wjk, Super-Kamiokande:2005lev,Super-Kamiokande:2014otb,Super-Kamiokande:2022egr} on the proton lifetime. An estimate of the rate of inclusive BNV $B$ decays shows that they are out of reach of direct experimental searches at current and future $B$-factories.

\section{Theoretical framework}
\label{sec:conv}

\subsection{BNV operators}
\label{sec:BNVops}
From the agnostic point of view on the origin of third-generation BNV adopted here, the Standard Model effective Lagrangian (SMEFT) including  
BNV dimension-six operators~\cite{Buchmuller:1985jz}
is the natural starting point for the discussion.
These operators are generated in the matching from an unknown fundamental theory at the scale $\LambdaBNV \gg m_W$.
We start from the Warsaw basis for SMEFT operators~\cite{Warsaw}. The baryon-number violating ones are
\begin{eqnarray}
\mathcal{Q}_{duu} &=& \varepsilon^{abc} [\widetilde{D}^a U^b]\;[\widetilde{U}^c E]\,,
\label{eq:duu}\\
\mathcal{Q}_{duq} &=& \varepsilon^{abc} \varepsilon_{jk}[\widetilde{D}^a U^b]\;[\widetilde{Q}_j^c L_k]\,,
\label{eq:duq}\\
\mathcal{Q}_{qqu} &=& \varepsilon^{abc} \varepsilon_{jk}[\widetilde{Q}_j^a Q_k^b]\;[\widetilde{U}^c E]\,,
\label{eq:qqu}\\
\mathcal{Q}_{qqq} &=& \varepsilon^{abc} \varepsilon_{jn} \varepsilon_{km}[\widetilde{Q}_j^a Q_k^b]\;[\widetilde{Q}_m^c L_n]\,,
\label{eq:qqq}
\end{eqnarray}
where spinor indices are contracted within the brackets, $a, b, c$ are colour indices ($\varepsilon^{123}=+1$), $j,k,m,n$ are SU(2)$_L$ doublets indices ($\varepsilon_{12} = +1$), and the fermion fields have an additional generation index (not shown) $p,r,s,t$, respectively, which stands for any of the three particle generations.
The left-handed doublets are denoted by $Q$ and $L$ for quarks and leptons, while the right-handed singlets are $U$, $D$ and $E$ for the up-, down-quark and charged lepton respectively. We can choose a specific generation basis for the SMEFT fields without loss of generality, by reabsorbing the five arbitrary rotation matrices for the fields $Q$, $U$, $D$, $L$ and $E$ into the SMEFT Wilson coefficients. We adopt the standard choice where the Yukawa coupling matrices $Y^u$ and $Y^e$ are diagonal and $Y^d_{rs} \propto V_{rt} [\text{diag}(m_d,m_s,m_b)]_{ts}$, with $V$ the Cabibbo-Kobayashi-Maskawa (CKM) matrix.
In this basis the transition from the weak eigenstates basis to the mass basis, after spontaneous symmetry breaking, requires only to perform the rotation 
$Q_{2p} = V_{pr}d_{Lr}$ of the second component of the SU(2)$_L$ doublet $Q$ in 
generation space.

The operators~\eqref{eq:duu}--\eqref{eq:qqq} are written in terms of charge-conjugated fields $\widetilde{\psi} \equiv \bar{\psi}^c$ which, in the Dirac representation, obey the following relations
\begin{equation}
\label{eq:chargeconj}
\psi^c = \mathcal{C}\bar{\psi}^T\,,  \qquad \bar{\psi}^c = -\psi^T \mathcal{C}^{-1}\,.
\end{equation}
The charge conjugation matrix $\mathcal{C} = i\gamma^2 \gamma^0$ satisfies
\begin{equation}
\label{eq:Cprop}
\mathcal{C}^\dagger = \mathcal{C}^{-1} = \mathcal{C}^T = -\mathcal{C}\,, \qquad \mathcal{C}\gamma_\mu^T \mathcal{C}^{-1} = -\gamma_\mu\,.
\end{equation}
Furthermore the operator structure can be simplified with the general relation
\begin{equation}
\label{eq:simprel}
\widetilde{\psi}_a P_{L,R}\, \psi_b = \widetilde{\psi}_b P_{L,R}\, \psi_a\,,
\end{equation}
where $a$ and $b$ stand for all the possible indices of the fields (colour, flavour, ...) and $P_{L(R)} = (1\mp \gamma^5)/2$ are the chiral projectors (not written explicitly in~\eqref{eq:duu}--\eqref{eq:qqq}).

The operators~\eqref{eq:duu}--\eqref{eq:qqq} arise from the matching of the unknown UV theory onto SMEFT at a scale $\mathcal{O}(\LambdaBNV)$.
The Wilson coefficients $C_{duu}^{prst}$, $C_{duq}^{prst}$, $C_{qqu}^{prst}$ and $C_{qqq}^{prst}$ are in principle generation-dependent and unknown. 
In order to study the effect of these BNV operators on low-energy physics such as $B$ meson and proton decays one has to evolve and match them onto the weak effective theory (WET) at the electroweak scale.
The renormalization group evolution of the BNV operators has been computed in~\cite{Alonso:2014zka}, their one-loop matching to the WET in~\cite{Dekens:2019ept}.
Since we are not interested in possible models at the high scale, our starting point for formulating the phenomenological assumption that BNV below the GUT scale occurs only in operators involving a third-generation quark will be the WET at the electroweak scale. At this scale spontaneous symmetry breaking takes place and the fields $u_p$, $d_p$ and $\ell_p$ are therefore written in the mass basis, which identifies what we 
mean by ``third generation''.

The WET dimension-6 operator basis\footnote{Table 4 in~\cite{Dekens:2019ept} shows many more BNV operators. However, most of them have vanishing Wilson coefficients at one-loop dimension-6 matching, hence are not listed here. The last equality in brackets refers to the notation of Table 4 in~\cite{Dekens:2019ept}.} is~\cite{Dekens:2019ept}
\begin{align}
\label{eq:WETBNVbasis}
    Q_{RR} &= \varepsilon^{abc} [\widetilde{d}^a_p P_R u^b_r]\;[\widetilde{u}^c_s P_R\ell_t] \quad\Bigl[= \mathcal{O}^{S,RR}_{duu}\Bigr]\,,\nonumber\\
    Q_{RL} &= \varepsilon^{abc} [\widetilde{d}^a_p P_R u^b_r]\;[\widetilde{u}^c_s P_L\ell_t] \quad\Bigl[= \mathcal{O}^{S,RL}_{duu}\Bigr]\,,\nonumber\\
    Q_{LR} &= \varepsilon^{abc} [\widetilde{d}^a_p P_L u^b_r]\;[\widetilde{u}^c_s P_R\ell_t] \quad\Bigl[= \mathcal{O}^{S,LR}_{duu}\Bigr]\,,\nonumber\\
    Q_{LL} &= \varepsilon^{abc} [\widetilde{d}^a_p P_L u^b_r]\;[\widetilde{u}^c_s P_L\ell_t] \quad\Bigl[= \mathcal{O}^{S,LL}_{duu}\Bigr]\,,\nonumber\\
    Q_{R\nu} &= \varepsilon^{abc} [\widetilde{d}^a_p P_R u^b_r]\;[\widetilde{d}^c_s P_L\nu_t] \quad\Bigl[= \mathcal{O}^{S,RL}_{dud}\Bigr]\,,\nonumber\\
    Q_{L\nu} &= \varepsilon^{abc} [\widetilde{d}^a_p P_L u^b_r]\;[\widetilde{d}^c_s P_L\nu_t] \quad\Bigl[= -\mathcal{O}^{S,LL}_{udd}\Bigr]\,,
\end{align}
where the generation indices $p,r,s,t$ on the left-hand side are omitted.
The BNV part of the WET Hamiltonian can be expressed compactly as
\begin{equation}
    \mathcal{H}_{\rm BNV} = \frac{1}{\LambdaBNV^2} \sum_{p,r,s,t}\;\sum_{X=L,R} \;\sum_{Y=L,R,\nu}C_{XY}^{prst} Q_{XY}^{prst} \,,
\end{equation}
where the generation sums do not include top quarks.
The dimensionless Wilson coefficients $C_{XY}^{prst}$ are linear combinations of the SMEFT Wilson coefficients $C_k^{prst}$ evaluated at the electroweak scale.

In this paper we investigate whether $\LambdaBNV \ll \Lambda_{\rm GUT} \sim \mathcal{O}(10^{15}~\text{GeV})$ is possible for the third generation, which in order to be viable needs anomalously small Wilson coefficients of operators with only light quarks.\footnote{The standard estimate for the scale of light-flavour BNV is rederived in 
Appendix~\ref{sec:lightBNV} for completeness.}
This therefore leads us to our working assumption on the Wilson coefficients in the WET
\begin{equation}
\label{eq:assumption}
C_{XY}^{prst}=0\,, \qquad \text{for} \qquad p,r,s \neq 3\,,
\end{equation}
for all operators $Q_{XY}$ from \eqref{eq:WETBNVbasis} above.
In other words we postulate the existence of BNV interactions at the low scale only if the operator contains (at least) one bottom quark field.
In the following we shall be looking for the least constrained BNV operators in order to obtain the largest possible branching fraction of a BNV $b$-hadron decay.

The case of operators including $b$ quarks and the $\tau$ lepton must be considered separately, since the proton decay directly into a $\tau$ is kinematically forbidden.\footnote{Operators with first-family quarks and a $\tau$ have already been constrained by considering proton decays mediated by a virtual $\tau$~\cite{Hou:2005iu}.}
These potentially interesting operators would mediate $\bar{B}\to \tau+X$ BNV decays and deserve a separate discussion, which is postponed to Section~\ref{sec:tauest}.
We will therefore mostly drop the lepton flavour index for simplicity, 
assuming it to be either an electron or a muon --- differences between $e$ and $\mu$ will enter only through mass effects and different experimental constraints.

Since the decays under investigation require a $b$ field and two first-family quark fields in the operator, we will focus on the BNV operators with this field content.
In the following subsections we will show that we can focus on only three operators, $Q_{RR}^{311}$, $Q_{RL}^{311}$ and $Q_{R\nu}^{311}$. 
Operators involving a $b$- and a second-generation quark might still be relevant for $B$ decays and will be briefly discussed in Section~\ref{sec:2ndfamily}.

\subsubsection{Operators with left-handed $b$ quark}

We need to identify the operators that allow for $\LambdaBNV \ll \Lambda_{\rm GUT}$ without having to require anomalously small dimensionless Wilson coefficients. 
In this way we will be able to bound the largest possible branching ratio for BNV $B$ decays in a model-independent approach. For this purpose, we select the Wilson coefficients of operators involving two light quarks and a $b$ quark that are least constrained by proton stability (for comparison, see~\eqref{eq:lightbound} in Appendix~\ref{sec:lightBNV}
for the constraint on light-flavour BNV coefficients).

We start by considering the WET operators from~\eqref{eq:WETBNVbasis} containing a left-handed $b$ quark. Due to our choice of the SMEFT basis for the fields discussed at the beginning of Section~\ref{sec:BNVops}, 
the tree-level matching to the WET simply consists of the rotation of left-handed down-type quarks to the mass basis through the CKM matrix.
This induces a correlation between the coefficients of operators with a left-handed $b$ quark and operators with only light quarks. 
As an example we consider the simplest case of the operator $Q_{R\nu}^{113}$, which is generated by $\mathcal{Q}_{duq}^{prs}$ in SMEFT.
The tree-level matching of the SMEFT operator to dimension-6 WET operators gives the Wilson coefficients
\begin{equation}
\label{eq:leftmatching}
    C_{RL}^{prs} = C_{duq}^{prs}\,,\qquad C_{R\nu}^{prs} = -C_{duq}^{prv} V_{vs}\,,
\end{equation}
where the sum over the flavour index $v$ is implied.

The relation~\eqref{eq:leftmatching} implies that the coefficient of $Q_{R\nu}^{113}$ (which contains a left-handed $b$ quark) depends on the same SMEFT building blocks as the coefficients of the light-BNV operators $Q_{RL}^{111}$, $Q_{R\nu}^{111}$ and $Q_{R\nu}^{112}$. 
More explicitly
\begin{align}
    C_{RL}^{111} &= C_{duq}^{111}\,,\nonumber\\
    C_{R\nu}^{111} &= - V_{ud} C_{duq}^{111} - V_{cd} C_{duq}^{112}  - V_{td} C_{duq}^{113}\,,\nonumber\\
    C_{R\nu}^{112} &= - V_{us} C_{duq}^{111} - V_{cs} C_{duq}^{112}  - V_{ts} C_{duq}^{113}\,,\nonumber\\
    C_{R\nu}^{113} &= - V_{ub} C_{duq}^{111} - V_{cb} C_{duq}^{112}  - V_{tb} C_{duq}^{113}\,,
\end{align}
which means that the condition~\eqref{eq:assumption}, $C_{R\nu}^{111}=C_{R\nu}^{112}=C_{RL}^{111}=0$, implies $C_{R\nu}^{113}=0$.
For the other WET operators with a left-handed $b$ quark the number of SMEFT coefficients  exceeds the number of equations. In principle, this allows large non-vanishing Wilson coefficients in the WET at the expense of highly fine-tuned relations between the underlying SMEFT Wilson coefficients, 
which are nevertheless not preserved by renormalization group evolution~\cite{Alonso:2014zka}. We conclude that the scale of left-handed 
third-generation BNV operators can be smaller than the one for light-flavoured operators only by a factor $\sqrt{|V_{ub}|}$ or $\sqrt{|V_{td}|}$.
Therefore we must focus on the operators with a right-handed $b$ quark. 

\subsubsection{Operators with right-handed $b$ quark}
\label{sec:rightb}

We are left with analyzing the WET operators composed of a right-handed $b$ quark and first family quarks:
\begin{align}
\label{eq:rightop}
   Q_{RR}^{311} =& \varepsilon^{abc}[\widetilde{b}^a P_R u^b]\,[\widetilde{\ell} P_R u^c]\,,\nonumber\\ 
   Q_{RL}^{311} =& \varepsilon^{abc}[\widetilde{b}^a P_R u^b]\,[\widetilde{\ell} P_L u^c]\,,\nonumber\\
   Q_{R\nu}^{311} =& \varepsilon^{abc}[\widetilde{b}^a P_R u^b]\,[\widetilde{\nu} P_L d^c]\,.
\end{align}
At tree-level, their Wilson coefficients 
are linked to the SMEFT ones by
\begin{align}
\label{eq:rightWC}
    C_{RR}^{311} &= C_{duu}^{311} \,,\nonumber \\
    C_{RL}^{311} &= C_{duq}^{311} \,,\nonumber \\
    C_{R\nu}^{311} &= -V_{ud} C_{duq}^{311} - V_{cd}C_{duq}^{312} - V_{td}C_{duq}^{313} \,.
\end{align}
Unlike the case of left-handed $b$ quark WET operators, 
the right-handed $b$ ones do not mix with light-BNV operators at dimension-6 level. 

\begin{figure}
    \centering
    \includegraphics[width=0.65\textwidth]{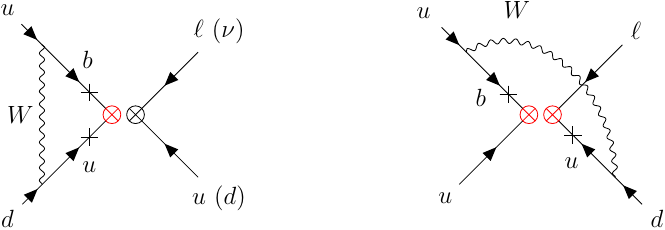}
    \caption{\small Two types of diagrams relevant for the dimension-8 matching of the operators $\mathcal{Q}_{duu}^{311}$ (left and right) and $\mathcal{Q}_{duq}^{311}$ (only left) into light-BNV operators. The red crosses $\red{\otimes}$ correspond to right-handed currents, while the black cross $\otimes$ stands for both right- and left-handed ones. Crosses on the propagators stand for mass insertions inducing the chirality flip 
    required to convert the right-handed quarks in the BNV operator to left-handed ones.}
    \label{fig:dim8match}
\end{figure}

However, the SMEFT operators $\mathcal{Q}_{duu}^{311}$ and $\mathcal{Q}_{duq}^{311}$, which source the left-hand sides of \eqref{eq:rightWC}, may also induce light-flavoured BNV operators at dimension-8 in the WET with suppressed 
matching coefficients through $W$-boson exchange. To estimate the amount of suppression, we consider one-loop electroweak corrections to SMEFT--WET matching as shown in Figure~\ref{fig:dim8match}. Since we want to match an operator containing the $b$ quark to an operator made only of light quarks, we have to consider flavour-changing weak interactions. We hence need a $W$ boson to couple to the right-handed $b$ quark from the operator to turn it into an external up quark, which requires a bottom-mass insertion 
to turn the $b$ quark left-handed. 
In order to close the loop, the $W$ must be attached to one of the remaining three fermions from the BNV operator.
In the case of $\mathcal{Q}_{duu}^{311}$ there are two possible diagrams (left and right in Figure~\ref{fig:dim8match}), where the $W$ is attached to one of the two remaining right-handed up quarks in the operator, since the attachment to the lepton is ruled out by charge conservation, thus incurring another chirality suppression proportional to $m_u$.
The case of $\mathcal{Q}_{duq}^{311}$ is even simpler because if the $W$ couples to the left-handed fermions of the operator, the loop integral will be odd in the loop momentum and vanish. Therefore for this operator only the diagram on the left of Figure~\ref{fig:dim8match} contributes. 
In all cases the internal right-handed quark legs will bring the factor $m_u m_b/m_W^2$ due to the double chirality flip needed by the weak interaction (shown as crosses on the propagators in the figure), implying that one is effectively matching the operators at dimension-8 in the WET counting.

Together with the CKM and SU(2) coupling factors, we obtain that the operators $\mathcal{Q}_{duq}^{311}$ and $\mathcal{Q}_{duu}^{311}$ match to light-flavoured BNV operators with coefficients $C_i^{\rm EW}$ given by
\begin{equation}
\label{eq:CEW}
    C_i^{\rm EW}(\mu) = V_{ud}V_{ub}^* \frac{m_u m_b}{4\pi^2v^2} \,F_i\Bigl(\ln \frac{\mu}{m_W}\Bigr) \approx 2\cdot 10^{-11} F_i\Bigl(\ln \frac{\mu}{m_W}\Bigr) \,.
\end{equation}
The $F_i$ are $\mathcal{O}(1)$ quantities and we expressed the SU(2) coupling $g_2^2 = 4m_W^2/v^2$ through the Higgs field vacuum expectation value $v=246.22~\text{GeV}$.
Using the results of appendix~\ref{sec:lightBNV}, this parametric suppression leads to constraints of the order
\begin{equation}
\label{eq:dim8bound}
    \frac{\LambdaBNV}{\sqrt{|C^{311}_{RY}|}} \gtrsim \sqrt{|C_i^{\rm EW}|}\cdot 10^{15}~\text{GeV} \gtrsim \mathcal{O}(10^{9})~\text{GeV}\,, \qquad Y=L,R,\nu\,,
\end{equation}
which implies that the effective scale of right-handed $b$-quark BNV can be 
five to six orders of magnitude below the GUT scale. In the next section we shall compare this result with the constraint from the direct computation of tree-level proton decays mediated by the same set of operators.

We now check that the assumption \eqref{eq:assumption} is stable under 
RGE mixing, that is, that the coefficients set to zero are not 
generated by the non-zero ones, when the scale $\mu\sim m_W$ is changed 
by, say, a factor of two. Since the Wilson coefficients of the right-handed $b$-quark WET BNV operators descend from the SMEFT coefficients $C^{311}_{duu}$ and $C^{311}_{duq}$ up to CKM 
factors, and since the mixing of the BNV SMEFT operators under renormalization is known~\cite{Alonso:2014zka}, 
we can check the magnitude of the mixing of $\mathcal{Q}_{duu}^{311}$, $\mathcal{Q}_{duq}^{311}$ into the other SMEFT operators which generate the strongly constrained operator $Q_{XY}^{111}$ after matching.
In practice we single out the presence of $C_{duu}^{311}$ and $C_{duq}^{311}$ in the RGE of the four first-family BNV Wilson coefficients $C^{111}_k$ ($k=duu,duq,qqu,qqq$), which are the dominant contribution to $C_{XY}^{111}$.
From~\cite{Alonso:2014zka} we obtain\footnote{Ref.~\cite{Alonso:2014zka} uses the opposite convention for the Yukawa matrices $Y^d|_{\text{this work}} \to {Y^d}^\dagger|_{\text{Alonso et al.}}$. Therefore in the convention of~\cite{Alonso:2014zka} the only non-diagonal matrices in flavour space are $Y^d$, ${Y^d}^\dagger$ and ${Y^d}^\dagger Y^d$.}
\begin{align}
\label{eq:SMEFT_RGE}
    \mu \frac{dC^{111}_{duu}}{d\mu} &= \dots\,,\nonumber\\
    \mu \frac{dC^{111}_{duq}}{d\mu} &= -V_{ud}V_{ub}^* \frac{m_d m_b}{8\pi^2 v^2}C_{duq}^{311}+ \dots\,,\nonumber\\
    \mu \frac{dC^{111}_{qqu}}{d\mu} &= V_{ub}^* \frac{m_\ell m_b}{8\pi^2 v^2}C_{duq}^{311}- 3V_{ub}^* \frac{m_u m_b}{8\pi^2 v^2}C_{duu}^{311}+\dots\,,\nonumber\\
    \mu \frac{dC^{111}_{qqq}}{d\mu} &= V_{ub}^* \frac{m_u m_b}{2\pi^2 v^2}C_{duq}^{311} +\dots\,,
\end{align}
where the dots denote the other Wilson coefficients we are not interested in. We notice that the parametric suppression is indeed the same as for the dimension-8 matching~\eqref{eq:CEW}. Therefore by a small change in the renormalization scale one generates at most a fraction $\mathcal{O}(10^{-11})$ of Wilson coefficients $C_{duq}^{311}$ and $C_{duu}^{311}$ for the light-flavoured operator. 
Hence the assumption \eqref{eq:assumption} is consistent with RG evolution, 
if the scale of the right-handed $b$-quark BNV operators is larger than 
\eqref{eq:dim8bound}. 

We therefore conclude that among the ones containing a $b$ quark, two first family quarks and a light lepton, the operators~\eqref{eq:rightop} are the least affected by the light-BNV operator constraints.

\subsubsection{Operators including second family quarks}
\label{sec:2ndfamily}

We briefly address the strategies for constraining operators including second family quarks and the right-handed $b$.
These operators could contribute to BNV $B_s$ meson  decays and $B$ decays to   final states with strangeness or charm.

We start by considering operators with a strange quark field.
There is only one operator with a right-handed $b$ quark and an $s$ quark, $Q_{R\nu}^{312} = \varepsilon^{abc}[\widetilde{b}^a P_R u^b]\,[\widetilde{\nu} P_L s^c]$.
This operator would mediate the decay $p \to K^+ \bar{\nu}$ by the same mechanism as $Q_{R\nu}^{311}$ does for $p \to \pi^+ \bar{\nu}$ in Figure~\ref{fig:diags1}.
However, the experimental constraint~\cite{Super-Kamiokande:2014otb} on $p \to K^+ \bar{\nu}$ partial lifetime is 15 times stronger than the one on $p \to \pi^+ \bar{\nu}$~\cite{Super-Kamiokande:2013rwg}.  
Since the theoretical calculation is identical to the one for $Q_{R\nu}^{311}$ presented below, we can discard this operator and focus only on the less constrained $Q_{R\nu}^{311}$, which therefore allows larger BNV $b$-hadron decay rates. 

On the other hand, for operators containing a bottom and a charm quark field, both, the $b$ and the $c$ have to be virtual, since the proton is too light to decay into charmed hadrons. Therefore the dominant constraint will come from 
electroweak mixing into the dimension-8 WET operators of the type discussed above, where one has to substitute $m_u V_{ud}\to m_c V_{cd}$. This implies an enhancement  by the factor $V_{cd}m_c/(V_{ud} m_u ) \sim 10^2$ in~\eqref{eq:CEW} which leads to the constraint  $\mathcal{O}(10^{10}~\text{GeV})$ on the scale of these operators. Thus $B$ decays into charmed hadrons are constrained more strongly than those to light hadrons. Hence, we shall focus on the latter below.


\subsection{Weak effective Hamiltonian}

The relevant part of the WET Hamiltonian for the proton decays of interest consists of the baryon-number violating part and the standard Hamiltonian for semi-leptonic and hadronic $b$ decays. The surviving BNV WET operators from~\eqref{eq:rightop} give the Hamiltonian
\begin{equation}
\mathcal{H}_{\rm BNV} = \frac{1}{\LambdaBNV^2} \Bigl(C_L Q_{RL}^{311} + C_R Q_{RR}^{311} + C_\nu Q_{R\nu}^{311}\Bigr) + \text{h.c.}\,.
\end{equation}
The Wilson coefficients~\eqref{eq:rightWC} have been renamed for convenience. 
The effective Hamiltonian for the weak semi-leptonic transition $b \to u \ell^- \bar{\nu}_\ell$ is
\begin{equation}
\mathcal{H}_{\rm sl} = 4\frac{G_F}{\sqrt{2}} V_{ub} Q_{\rm sl} + {\rm h.c.}\,,
\end{equation}
with
\begin{equation}
Q_{\rm sl}^\dagger = [\bar{b}\gamma^\mu P_L u]\; [\bar{\nu}_\ell \gamma_\mu P_L \ell]\,.
\end{equation}
The one for the 
hadronic  $b \to u \bar{u}d$ transition reads 
\begin{equation}
\mathcal{H}_W = 4\frac{G_F}{\sqrt{2}} V_{ub}V_{ud}^*\left(C_1 Q_1 + C_2 Q_2 \right) + \text{h.c.}\,,
\end{equation}
with the two operators 
\begin{align}
Q_1^\dagger &= [\bar{b}\gamma^\mu P_L T^A u]\,[\bar{u}\gamma_\mu P_L T^A d]\,,\nonumber\\
Q_2^\dagger &= [\bar{b}\gamma^\mu P_L u]\,[\bar{u}\gamma_\mu P_L d]\,.
\end{align}
The quark colour and spinor indices are summed within the squared brackets and $T^A$ are the SU(3) colour generators in the fundamental representation. We focus on 
the dominant charged-current operators, and neglect the loop-suppressed 
penguin operators.

\subsection{Contact six-fermion interaction}

The WET operators are evolved from the electroweak scale to $m_b$, at which the heavy virtual $b$ quark in the diagrams of Figure~\ref{fig:diags1} is integrated out, resulting in the local six-fermion operator
\begin{equation}
\label{eq:Onusl}
\mathcal{O}_{\nu, \text{sl}} = \varepsilon^{abc} [\widetilde{u}^a \gamma^\mu P_L u^b]\; [\widetilde{\nu} P_L d^c]\; [\bar{\nu}_\ell \gamma_\mu P_L \ell]\,,
\end{equation}
for $p \to \ell^+ \nu_\ell \bar{\nu}$, and
\begin{align}
\mathcal{O}_{\nu,1} &= \varepsilon^{abc}[\widetilde{u}^a \gamma^\mu P_L T^A_{bi}  u^i]\; [\widetilde{\nu}\,  P_L d^c]\; [\bar{u}^f \gamma_\mu P_L T^A_{fj} d^j]\,,\nonumber\\
\mathcal{O}_{\nu,2} &= \varepsilon^{abc} [\widetilde{u}^a \gamma^\mu P_L u^b]\; [\widetilde{\nu}\,  P_L d^c]\; [\bar{u}^f \gamma_\mu P_L d^f]\,,\nonumber\\
\mathcal{O}_{X,1} &= \varepsilon^{abc}[\widetilde{u}^a \gamma^\mu P_L T^A_{bi}  u^i]\; [\widetilde{\ell}\,  P_X u^c]\; [\bar{u}^f \gamma_\mu P_L T^A_{fj} d^j]\,,\nonumber\\
\mathcal{O}_{X,2} &= \varepsilon^{abc} [\widetilde{u}^a \gamma^\mu P_L u^b]\; [\widetilde{\ell}\,  P_X u^c]\; [\bar{u}^f \gamma_\mu P_L d^f]\,,
\end{align}
for $p \to \pi^+ \bar{\nu}$ and $p \to \pi^0 \ell^+$.
For all cases the tree-level matching coefficient is $1/m_b$,
and the subscripts refer to the BNV and weak currents, respectively, and $X=L, R$. 
For the colour-singlet current with two up quarks only the vectorial part 
contributes. The axial-vector part vanishes due to the identity 
\begin{equation}
\label{eq:symprop}
\varepsilon^{abc} [{u^a}^T \mathcal{C} \gamma^\mu P_L u^b] = \varepsilon^{abc} [{u^a}^T \mathcal{C} \gamma^\mu P_L u^b]^T = -\varepsilon^{abc} [{u^b}^T P_L {\gamma^\mu}^T\mathcal{C}^T u^a] = \varepsilon^{abc}[{u^a}^T \mathcal{C} \gamma^\mu P_R u^b]\,.
\end{equation}
We can therefore write the Hamiltonian relevant for the processes of interest as
\begin{equation}
    \mathcal{H}_{\rm 6f} = \mathcal{H}_{p \to \ell^+ \nu_\ell \bar{\nu}} + \mathcal{H}_{p \to \pi^+ \bar{\nu}} +\mathcal{H}_{p \to \pi^0\ell^+} \,,
\end{equation}
with 
\begin{align}
    \mathcal{H}_{p \to \ell^+ \nu_\ell \bar{\nu}} &= -2\sqrt{2}\,\frac{G_F C_\nu V_{ub}^*}{m_b \LambdaBNV^2} \,\mathcal{O}_{\nu,\text{sl}}  + \text{h.c.}\,,\nonumber\\
    \mathcal{H}_{p \to \pi^+ \bar{\nu}} &= -2\sqrt{2}\,\frac{G_F C_\nu V_{ub}^* V_{ud}}{m_b \LambdaBNV^2}\Bigl(C_1 \mathcal{O}_{\nu,1} + C_2 \mathcal{O}_{\nu,2} \Bigr) + \text{h.c.}\,,\nonumber\\
    \mathcal{H}_{p \to \pi^0\ell^+} &= -2\sqrt{2}\,\frac{G_F V_{ub}^*V_{ud}}{m_b \LambdaBNV^2}\sum_{X=L,R} C_X \Bigl(C_1 \mathcal{O}_{X,1} + C_2 \mathcal{O}_{X,2} \Bigr) + \text{h.c.}\,.
\end{align}


\section{Proton decay rate}

The decay rate in the rest frame of the decaying proton is obtained with the standard formula
\begin{equation}
\label{eq:genform}
    \Gamma(p\to f) = \frac{1}{2m_p} \int d\Pi_{\rm LIPS} \,\frac{1}{2}\sum_{\rm spins}|\langle f|\mathcal{H}_{\rm 6f}| p \rangle|^2\,,
\end{equation}
where $f$ denotes the final state, and the factor $1/2$ is to average over the proton spin.
The Lorentz-invariant phase space reads 
\begin{equation}
d\Pi_{\rm LIPS} = (2\pi)^4 \delta^4\Bigl(p_{in} - \sum_{\text{final }j}p_j\Bigr) \prod_{\text{final\,} j} \frac{d^4p_j}{(2\pi)^3}\delta(p_j^2-m_j^2)\theta(p^0_j)\,.
\end{equation}
For the hadronic matrix elements we use the expressions 
\begin{align}
\label{eq:projectors}
\delta^{ab}\Pi_{\beta\alpha}(p) &\equiv \langle \pi^+(p) | \bar{u}^a_\alpha\; d^b_\beta|0\rangle = \frac{i}{4N_c}\delta^{ab}f_\pi (\slashed{p}\gamma^5 -\mu_\pi \gamma^5)_{\beta\alpha}\,,\nonumber\\
G_{\alpha\beta\gamma}(p) &\equiv \langle 0| \varepsilon^{abc} \widetilde{u}^a_\alpha\, u^b_\beta \, d^c_\gamma \, |p(p)\rangle = -\frac{f_p}{4}\biggl(\slashed{p}_{\beta\alpha}[\gamma^5 u_p(p)]_\gamma + i p^\nu [\sigma_{\rho\nu}]_{\beta\alpha} [\gamma^\rho \gamma^5 u_p(p)]_\gamma  \biggr) \nonumber\\
&+ \frac{m_p}{16}(\lambda_1-f_p)[\gamma_\rho]_{\beta\alpha}[\gamma^\rho \gamma^5 u_p(p)]_\gamma +  \frac{m_p}{96}(\lambda_2 -6f_p)[\sigma_{\rho\sigma}]_{\beta\alpha} [\sigma^{\rho \sigma}\gamma^5 u_p(p)]_\gamma \,,
\end{align}
where $u_p(p)$ denotes the proton spinor, the spinor indices $\alpha$, $\beta$, $\gamma$ 
are uncontracted, and the fields are evaluated at $x=0$. 
The derivation is relegated to Appendix~\ref{sec:appproj}.
The parameters $f_p$, $\lambda_1$ and $\lambda_2$ (real, mass dimension 2) are the proton decay constants, with 
numerical values summarized in Appendix~\ref{sec:inputs}, together with the definition of $\mu_\pi$. $N_c=3$ denotes the number of colours.
For convenience, we also use
\begin{equation}
\Omega_p \equiv \frac{1}{4}(\lambda_1-f_p)\,.
\end{equation}

\subsection{Leptonic decay: $p\to \ell^+ \nu_\ell \bar{\nu}$}

\subsubsection{Decay rate} 

For the leptonic three-body decay the general formula~\eqref{eq:genform} reduces to
\begin{equation}
\label{eq:leptdec}
\Gamma(p\to \ell^+ \nu_\ell \bar{\nu}) = \frac{4 G_F^2|V_{ub}|^2 |C_\nu|^2}{m_p m_b^2\LambdaBNV^4}\int d\Pi_{\rm LIPS}\,\frac{1}{2}\sum_{\rm spins}|\langle \ell^+ \nu_\ell \bar{\nu} |\mathcal{O}_{\nu,\text{sl}}| p \rangle|^2 \,.
\end{equation}
We compute the matrix element of the six-fermion operator by splitting it in hadronic and leptonic contributions as
\begin{align}
\mathcal{O}_{\nu,\text{sl}} =  \frac{1}{2}\,[\bar{\nu}_\ell\gamma_\mu P_L \ell]\;[\widetilde{\nu} P_L]_\alpha[{\mathcal{O}}_q^\mu]_\alpha \,, \qquad \mathcal{O}_{\nu,\text{sl}}^\dagger = \frac{1}{2}\,[\bar{\ell} \gamma_\nu P_L \nu_\ell]\;[\bar{\mathcal{O}}_q^\nu]_\beta\; [P_R \nu^c]_\beta\; \,,
\end{align}
where we used the property~\eqref{eq:symprop} to write
\begin{align}
\mathcal{O}_q^\mu = \varepsilon^{abg} [\widetilde{u}^a \gamma^\mu u^b]\,d^g \,,\qquad \bar{\mathcal{O}}_q^\nu = \varepsilon^{abg} [\bar{u}^b \gamma^\nu (u^c)^a]\, \bar{d}^g\,.
\end{align}

With these definitions the matrix element squared is
\begin{equation}
\label{eq:matel}
\frac{1}{2}\sum_{\rm spins}|\langle \ell^+ \nu_\ell  \bar{\nu}|\mathcal{O}_{\nu,\text{sl}}| p \rangle|^2 = \frac{1}{4}\,L_{\mu\nu} \sum_{p~{\rm spin}}\text{Tr}\Bigl[\langle 0|\mathcal{O}_q^\mu |p \rangle \langle p|\bar{\mathcal{O}}_q^\nu |0\rangle \slashed{q} P_L \Bigr] \,.
\end{equation}
Here $q$ denotes the momentum of the anti-neutrino $\bar{\nu}$, and the leptonic tensor from the lepton pair $\nu_\ell (p_n) \ell^+ (p_\ell)$ is given by
\begin{equation}
\label{eq:leptens}
L^{\mu\nu} = p_\ell^\mu p_n^\nu - p_\ell \cdot p_n g^{\mu\nu} + p_\ell^\nu p_n^\mu + i \varepsilon^{\mu\nu\alpha\beta}{p_\ell}_\alpha {p_n}_\beta\,. 
\end{equation}
Employing \eqref{eq:projectors}, the hadronic matrix element is given by 
\begin{equation}
\label{eq:hadpar}
\langle 0 |\mathcal{O}_q^\mu |p(p)\rangle = - f_p p^\mu [\gamma^5 u_p(p)] + m_p\Omega_p[\gamma^\mu \gamma^5 u_p(p)]\,. 
\end{equation}
Altogether, we find for  the matrix element~\eqref{eq:matel} of the six-fermion operator 
\begin{align}
\label{eq:6fmatel}
\frac{1}{2}\sum_{\rm spins}|\langle \ell^+(p_\ell) \nu_\ell(p_n)  \bar{\nu}(q)|\mathcal{O}_{\nu,\text{sl}}| p(p) \rangle|^2 &= m_p^6 \,\Bigl[\hat{\mathcal{M}}_{ff}\, f_p^2 + \hat{\mathcal{M}}_{f\Omega} \,f_p \Omega_p + \hat{\mathcal{M}}_{\Omega\Omega}\,\Omega^2_p \Bigr]\,,
\end{align}
with coefficients
\begin{align}
\hat{\mathcal{M}}_{ff} =& \;\frac{1}{2}\hat{E}_q(2 \hat{E}_\ell \hat{E}_n - \hat{p}_\ell \cdot \hat{p}_n)\,,\nonumber\\
\hat{\mathcal{M}}_{f\Omega} =& \;\hat{E}_\ell \hat{q}\cdot \hat{p}_n + \hat{E}_n \hat{q}\cdot \hat{p}_\ell - \hat{E}_q \hat{p}_\ell \cdot \hat{p}_n \,,\nonumber\\
\hat{\mathcal{M}}_{\Omega\Omega} =& \;2\hat{E}_n \hat{q}\cdot \hat{p}_\ell\,,
\end{align}
where we used $p = (m_p,0,0,0)$, and hatted expressions refer to the quantities 
normalized by appropriate powers of the proton mass: $\hat{E}_\ell = E_\ell/m_p$, $\hat{E}_n = E_n/m_p$ and $\hat{E}_q = E_q/m_p$ for the charged lepton, neutrino and anti-neutrino energies, respectively.

The phase-space integration can be reduced to 
\begin{eqnarray}
\int d\Pi_{\rm LIPS} &=& \frac{1}{4(2\pi)^3}\int_0^\infty dE_q \int_{m_\ell}^\infty dE_\ell \,\theta\Bigl(m_p - E_\ell -E_q -\bigl|E_q-\sqrt{E_\ell^2-m_\ell^2}\bigl|\Bigr)
\nonumber\\
&&\times\,\theta\Bigl(2E_q + E_\ell + \sqrt{E_\ell^2-m_\ell^2}-m_p\Bigr)\,,
\end{eqnarray}
resulting in
\begin{eqnarray}
\Gamma(p \to \ell^+ \nu_\ell \bar{\nu}) &=& |V_{ub}|^2 |C_\nu|^2 \,\frac{G_F^2 m_p^7}{7680 \pi^3 m_b^2 \LambdaBNV^4}\nonumber\\
&&\hspace*{-2cm}\times\,\Bigl[(1- \hat{m}_\ell^2)^5 f_p^2 + \frac{5}{8}(1-8\hat{m}_\ell^2+8\hat{m}_\ell^6-\hat{m}_\ell^8-24\hat{m}_\ell^4 \ln \hat{m}_\ell)(\lambda_1^2-f_p^2)\Bigr]\,,
\end{eqnarray}
with the normalized lepton mass $\hat{m}_\ell = m_\ell/m_p$. 

For the proton decay constants we use the numerical values evolved to 1~GeV reported in Appendix~\ref{sec:inputs}.
With the other numerical inputs and experimental limits on the partial lifetimes $\tau_{p\to e^+\nu\nu}$, $\tau_{p\to \mu^+\nu\nu}$ from Table~\ref{tab:input} in Appendix~\ref{sec:inputs} we then find the 
following lower limits on the scale of BNV:
\begin{align}
\label{eq:leptbounds}
\frac{\LambdaBNV}{\sqrt{|C_\nu|}}\biggl|_{p \to e^+ \nu_e \bar{\nu}} &> 6.59\cdot 10^9~\text{GeV}\,,\nonumber\\
\frac{\LambdaBNV}{\sqrt{|C_\nu|}}\biggl|_{p \to \mu^+ \nu_\mu \bar{\nu}} &> 6.86\cdot 10^9~\text{GeV}\,.
\end{align}

\subsubsection{Lepton energy spectrum}

\begin{figure}
\centering
\includegraphics[width=0.6\textwidth]{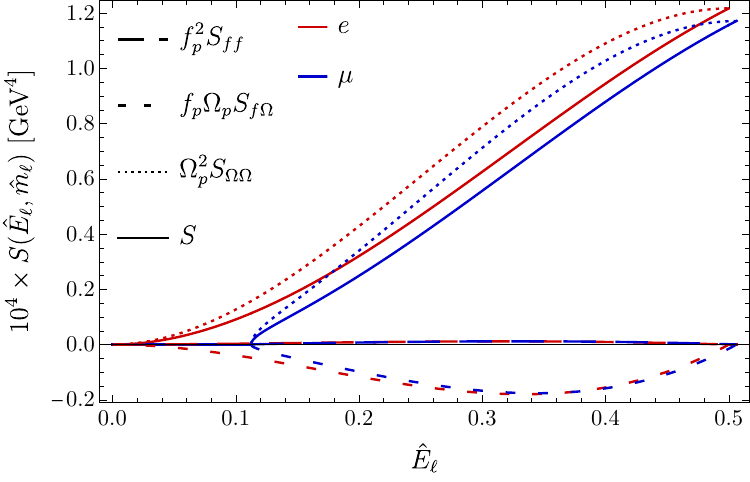}
\caption{\small Lepton energy spectrum (solid) for an electron (red) or a muon (blue) in the final state. Different dashed lines represent the three separate contributions of \eqref{eq:speccont} weighted by their prefactors.}
\label{fig:leptspec2}
\end{figure}

The spectrum in the lepton energy can be obtained by integrating only over the anti-neutrino energy $E_q$,
\begin{align}
\frac{d\Gamma}{d\hat{E}_\ell} =& \, |V_{ub}|^2 |C_\nu|^2\frac{G_F^2m_p^7}{(2\pi)^3\LambdaBNV^4 m_b^2}\int_0^\infty d\hat{E}_q \Bigl[\hat{\mathcal{M}}_{ff}\, f_p^2 + \hat{\mathcal{M}}_{f\Omega} \,f_p \Omega_p + \hat{\mathcal{M}}_{\Omega\Omega}\,\Omega^2_p \Bigr] \nonumber\\
&\times\theta\Bigl(1 - \hat{E}_\ell -\hat{E}_q -\bigl|\hat{E}_q-\sqrt{\hat{E}_\ell^2-\hat{m}_\ell^2}\bigl|\Bigr)\theta\Bigl(2\hat{E}_q + \hat{E}_\ell + \sqrt{\hat{E}_\ell^2-\hat{m}_\ell^2}-1\Bigr)\nonumber\\
=& \,|V_{ub}|^2 |C_\nu|^2 \frac{G_F^2m_p^7}{192\pi^3\LambdaBNV^4 m_b^2} \,\theta(\hat{E}_\ell-\hat{m}_\ell)\,\theta\left(\frac{1+\hat{m}_\ell^2}{2}-\hat{E}_\ell \right)\nonumber\\
&\times\Bigl[S_{ff}(\hat{E}_\ell,\hat{m}_\ell)f_p^2 + S_{f\Omega}(\hat{E}_\ell,\hat{m}_\ell)f_p \Omega_p + S_{\Omega\Omega}(\hat{E}_\ell,\hat{m}_\ell)\Omega_p^2 \Bigr] \,,
\end{align}
with functions $S_{ij}(\hat{E}_\ell,\hat{m}_\ell)$ given by
\begin{align}
\label{eq:speccont}
S_{ff}(\hat{E}_\ell,\hat{m}_\ell) =& \sqrt{\hat{E}_\ell^2-\hat{m}_\ell^2} \left(4 \hat{E}_\ell^3-8 \hat{E}_\ell^2+\hat{E}_\ell \left(3-\hat{m}_\ell^2\right)+2 \hat{m}_\ell^2\right)\,,\nonumber\\
S_{f\Omega}(\hat{E}_\ell,\hat{m}_\ell) =& 12\hat{E}_\ell \sqrt{\hat{E}_\ell^2-\hat{m}_\ell^2}(1-2\hat{E}_\ell+\hat{m}_\ell^2)\,,\nonumber\\
S_{\Omega\Omega}(\hat{E}_\ell,\hat{m}_\ell) =& 4\sqrt{\hat{E}_\ell^2-\hat{m}_\ell^2}\left(3\hat{E}_\ell(1+\hat{m}_\ell^2)-4\hat{E}_\ell^2-2\hat{m}_\ell^2\right)\,.
\end{align}
We display the lepton energy spectra 
\begin{equation}
S(\hat{E}_\ell,\hat{m}_\ell) \equiv \Bigl[S_{ff}(\hat{E}_\ell,\hat{m}_\ell)f_p^2 + S_{f\Omega}(\hat{E}_\ell,\hat{m}_\ell)f_p \Omega_p + S_{\Omega\Omega}(\hat{E}_\ell,\hat{m}_\ell)\Omega_p^2 \Bigr]\,,
\end{equation}
for the electron and muon by solid lines in Figure~\ref{fig:leptspec2}, and the 
three terms separately in dashed. It is apparent that the term proportional to $\Omega_p^2$ dominates over the others.

\subsubsection{Estimate of $p \to \ell^+ \nu_\ell \bar{\nu}_\tau$ mediated by a virtual $\tau$}
\label{sec:tauest}

\begin{figure}
    \centering
    \includegraphics[width=0.35\textwidth]{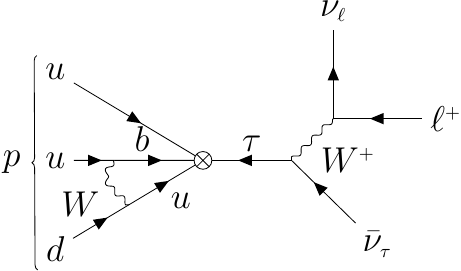}
    \caption{\small Proton decay $p \to \ell^+ \nu_\ell \bar{\nu}_\tau$ mediated by a $Q_{RX}^{3113}$ operator and a virtual $\tau$ lepton. The $W$ propagators and the loop are effectively integrated out in the WET, but displayed for clarity.}
    \label{fig:diagtau}
\end{figure}

In order to estimate the maximally allowed rates for $B$ decays into 
a final state with a $\tau$ lepton, we must consider constraints from proton decay on the BNV operators \eqref{eq:rightop} involving $b$ and $\tau$ simultaneously. They require a  separate treatment, because the proton cannot decay kinematically into a $\tau$ lepton. 
In the following we provide simple estimates for such operators.

An efficient way to estimate the constraint on operators involving a $b$ and a $\tau$ considers proton decay through a virtual $\tau$-lepton 
by a light-quark-flavoured BNV operator with a $\tau$, which in turn is 
generated from a third-quark generation SMEFT BNV operator through electroweak matching to dimension-8 WET operators discussed in Section~\ref{sec:rightb}. The process is illustrated in Figure~\ref{fig:diagtau}.
In other words, we effectively consider a light-flavoured BNV operator with a $\tau$ and coefficient given by the product of $C_{RX}^{3113}$ and 
the matching coefficient~\eqref{eq:CEW}. One then computes the decay rate of $p \to \ell^+ \nu_\ell \bar{\nu}_\tau$ by integrating out the virtual $\tau$ propagator, matching to a local six-fermion operator.
The resulting six-fermion operator has a Dirac structure different from  $\mathcal{O}_{\nu,\text{sl}}$ in~\eqref{eq:Onusl}, since the diagrams in Figure~\ref{fig:diagtau} and Figure~\ref{fig:diags1} (left) 
have a different ordering of the weak and BNV vertex. 
With the exception of the hadronic matrix element the computation follows the same steps as above for the fully leptonic decay, with the virtual $b$ propagator $1/m_b$ substituted by the virtual $\tau$ propagator $m_\tau/(m_\tau^2-m_p^2)$.\footnote{We keep the proton mass since $m_p \ll m_\tau$ is a poor numerical approximation.}

Therefore we estimate the constraint on $C_{RX}^{3113}$ by multiplying the bound~\eqref{eq:leptbounds} by the dimension-8 matching factor~\eqref{eq:CEW} (except for $V_{ub}$ which is already present in the weak effective Hamiltonian for the leptonic decay) and by the propagator ratio 
$m_b m_\tau/(m_\tau^2-m_p^2)$. To account for the uncomputed $\mathcal{O}(1)$ factor in $C_i^{\rm EW}$ and the different hadronic matrix element we 
allow for an uncertainty factor $(0.2\div 4)$.
This results in
\begin{equation}
    \frac{\LambdaBNV}{\sqrt{|C_{RX}^{3113}|}} \approx \left((0.2\div 4)\,\frac{|V_{ud}|m_b^2 m_u m_\tau}{4\pi^2 v^2 (m_\tau^2-m_p^2)}\right)^{\!1/2}\,\frac{\LambdaBNV}{\sqrt{|C_\nu|}}\biggl|_{p \to \ell^+ \nu_\ell \bar{\nu}_\tau} \gtrsim \,(0.4\div 1.8) \cdot 10^6~\text{GeV}\,,
    \label{eq:tauoperator}
\end{equation}
where $m_\tau=1.777~\text{GeV}$, which we shall use below to estimate 
the possible magnitude of BNV $B$ decays into $\tau$ final states.

\subsection{Two-body decay $p \to \pi^+ \bar{\nu}$}
\label{sec:ptopipnu}

\begin{figure}[t]
\centering
\includegraphics[width=0.7\textwidth]{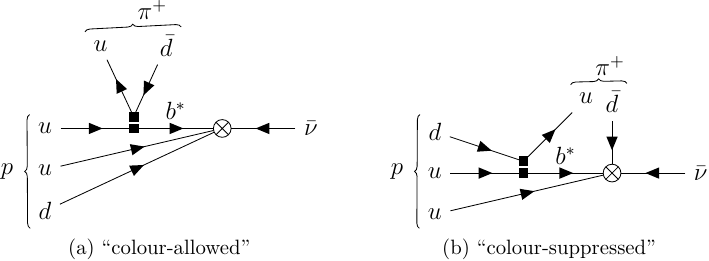}
\caption{\small Factorizable tree topologies contributing to $p \to \pi^+ \bar{\nu}$. 
The virtual $b$-quark propagator is integrated out but displayed for clarity. For $p \to \pi^0 \ell^+$ the colour-allowed topology is absent.}
\label{fig:diags2}
\end{figure}

We now turn to the two-body proton decays into a pion and a lepton 
(see Figure~\ref{fig:diags2} for the relevant Feynman diagrams). 
To calculate the hadronic 
matrix element, we adopt the naive factorization 
assumption that neglects strong interaction effects connecting the pion to the rest of the transition. Unlike 
the case of heavy hadron decay, for protons the factorization assumption is only an $\mathcal{O}(1)$ 
approximation. 

\begin{figure}[t]
\centering
\includegraphics[width=0.7\textwidth]{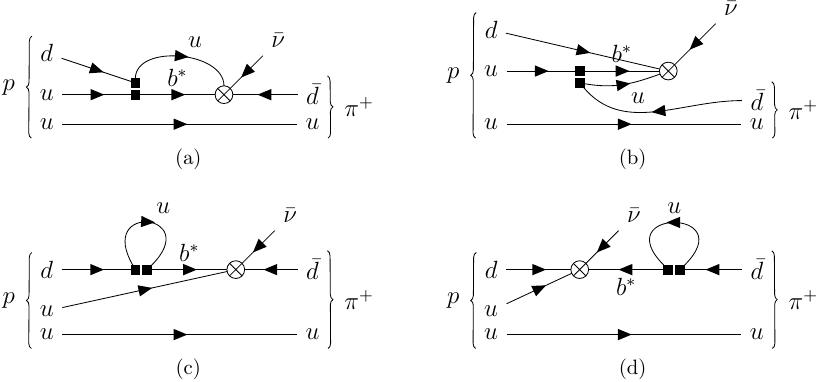}
\caption{\small Non-factorizable contributions to $p\to \pi^+ \bar{\nu}$. The virtual $b$-quark propagator is integrated out but displayed for clarity. Identical topologies exist for the $p \to \pi^0 \ell^+$ with appropriate substitutions.}
\label{fig:diags3}
\end{figure}

In addition to the two tree-level topologies of Figure~\ref{fig:diags2}, there are four ``non-factorizable'' loop diagrams 
(shown in Figure~\ref{fig:diags3}) 
to which the 
naive factorization approximation cannot be applied.  
The hadronic physics is contained in the baryon-number 
violating proton to pion form factor.
However, when computing the loop integrals, diagrams (c) and (d) are identically zero because the loop integrand is odd in the loop momentum.
On the other hand the first two are proportional to the up-quark mass, and parametrically of the same order of the dimension-8 contributions derived in~\eqref{eq:CEW}. We therefore consider the two factorizable diagrams of Figure~\ref{fig:diags2}, since the non-factorizable ones would give at most a bound of the same order as~\eqref{eq:dim8bound}.

From~\eqref{eq:genform} the decay rate is given by
\begin{equation}
\label{eq:2bodydec}
\Gamma(p\to \pi^+\bar{\nu}) = \frac{4G_F^2|V_{ub}|^2|V_{ud}|^2 |C_\nu|^2}{m_p m_b^2 \LambdaBNV^4}\int d\Pi_{\rm LIPS}\,\frac{1}{2}\sum_{\rm spins}|\langle \pi^+ \bar{\nu} |C_1\mathcal{O}_{\nu,1}+C_2\mathcal{O}_{\nu,2}| p \rangle|^2 \,.
\end{equation}
 By using~\eqref{eq:projectors} we can compute simultaneously the two diagrams contributing to the matrix element of the operator $\mathcal{O}_{\nu,2}$ 
as
\begin{eqnarray}
\langle \pi^+(p_\pi) \bar{\nu}(q)|\mathcal{O}_{\nu,2} |p(p)\rangle &=& \frac{1}{2}\,[v^T(q) \mathcal{C}P_L]_\gamma \Bigl\{N_c\,G_{\alpha\beta\gamma}(p)[\gamma^\mu]_{\alpha\beta} \text{Tr}[\Pi(p_\pi)\gamma_\mu P_L]\nonumber\\
&&\hspace*{-2.6cm}-\,G_{\alpha\beta\rho}(p)[\gamma^\mu]_{\alpha\beta} [\Pi(p_\pi)\gamma_\mu P_L]_{\gamma\rho} \Bigr\}\nonumber\\
&& \hspace*{-3cm}=\, \frac{i}{8N_c}m_p^2 f_\pi \Bigl\{(2N_c \hat{E}_\pi -1)f_p +(2N_c-4)\Omega_p\Bigr\}[v^T(q) \mathcal{C}P_L u_p(p)] \,,\qquad
\end{eqnarray}
where $\hat{E}_\pi=E_\pi/m_p$ is the normalized pion energy in the proton rest frame.
In the case of the matrix element of $\mathcal{O}_{\nu,1}$ the ``colour-allowed'' diagram vanishes due to the colour algebra, and we are left only with the ``colour-suppressed'' topology, resulting in
\begin{eqnarray}
\langle \pi^+(p_\pi) \bar{\nu}(q)|\mathcal{O}_{\nu,1} |p(p)\rangle &=& \frac{N_c+1}{4N_c}[v^T(q) \mathcal{C}P_L]_\gamma \Bigl\{G_{\alpha\beta\rho}(p)[\gamma^\mu(1-\gamma^5)]_{\alpha\beta} [\Pi(p_\pi)\gamma_\mu P_L]_{\gamma\rho} \Bigr\}\nonumber\\
&=& \frac{N_c+1}{4N_c^2} \frac{i}{4}m_p^2 f_\pi^2 (f_p+4\Omega_p)[v^T(q) \mathcal{C}P_L u_p(p)]\,.
\end{eqnarray}
When squaring the matrix element and summing over the spins we use
\begin{align}
\sum_{\rm spins} |v^T(q) \mathcal{C}P_L u_p(p)|^2 &= \sum_{\rm spins} \text{Tr}[\mathcal{C}(v(q)\bar{v}(q))^T\mathcal{C} P_L u_p(p)\bar{u}_p(p) P_R] \nonumber\\
&= \text{Tr}[\slashed{q} P_L (\slashed{p} + m_p) P_R] = 2m_p^2 \hat{E}_q = m_p^2(1 - \hat{m}_\pi^2)\,.
\end{align}
Therefore the result for the matrix element square reads
\begin{align}
\frac{1}{2}\sum_{\rm spins}|\langle \pi^+ \bar{\nu}|C_1\mathcal{O}_{\nu,1}+C_2\mathcal{O}_{\nu,2} |p\rangle|^2 =& \,
\frac{m_p^6}{64N_c^2}f_\pi^2 \hat{E}_q\Bigl\{C_2\left[(2N_c \hat{E}_\pi -1)f_p + (2N_c-4)\Omega_p\right] \nonumber\\
&+C_1\frac{N_c+1}{2N_c}(f_p+4\Omega_p)\Bigr\}^2\,.
\end{align}
Including the two-body phase-space integral
\begin{equation}
\int d\Pi_{\rm LIPS} = \frac{1}{8\pi}(1-\hat{m}_\pi^2)\,,
\end{equation}
we find from~\eqref{eq:2bodydec}:
\begin{eqnarray}
\Gamma(p \to \pi^+ \bar{\nu}) &=& |V_{ud}|^2|V_{ub}|^2 |C_\nu|^2\frac{G_F^2 m_p^5 f_\pi^2}{1024\pi m_b^2 \Lambda_{\rm  NP}^4}(1-\hat{m}_\pi^2)^2\nonumber\\
&&\times\left[\left((1 + 2 \hat{m}_\pi^2)f_p+\frac{\lambda_1}{3}\right)C_2 + \frac{4}{9}\lambda_1 C_1\right]^2 \,.
\end{eqnarray}
Evaluating the Wilson coefficients $C_1$ and $C_2$ at the scale 1~GeV with next-to-next-to-leading order  running~\cite{Gorbahn:2004my}, as reported in Appendix~\ref{sec:inputs}, this equation translates into the bound
\begin{equation}
\frac{\LambdaBNV}{\sqrt{|C_\nu|}}\biggl|_{p \to \pi^+ \bar{\nu}} > 3.34 \cdot 10^9~\text{GeV}
\end{equation}
on the third-generation BNV scale.
By comparison, the bound~\eqref{eq:leptbounds} obtained from the fully leptonic channel, which is sensitive to the same WET operator $Q_{R\nu}^{311}$, turns out to be somewhat stronger.


\subsection{Two-body decay $p \to \pi^0 \ell^+$}
\label{sec:ptopi0ell}

For this process we can use the most stringent experimental constraints on proton decays~\cite{Super-Kamiokande:2020wjk}.
The decay rate reads
\begin{equation}
\label{eq:Gammappi0ell}
\Gamma(p\to \pi^0\ell^+) = \frac{4G_F^2|V_{ub}|^2|V_{ud}|^2}{m_p m_b^2 \LambdaBNV^4}\int d\Pi_{\rm LIPS}\,\frac{1}{2}\sum_{\rm spins}\Bigl|\sum_{X=L,R}\sum_{i=1,2} C_X C_i \langle \pi^0 \ell^+ |\mathcal{O}_{X,i}| p \rangle\Bigr|^2 \,,
\end{equation}
where here the phase-space factor with a massive lepton is
\begin{equation}
    \int d\Pi_{\rm LIPS} = \frac{1}{8\pi} \sqrt{(1-\hat{m}_\pi^2)^2 -2\hat{m}_\ell^2 (1+\hat{m}_\pi^2)+\hat{m}_\ell^4}\,.
\end{equation}
The matrix elements are given by
\begin{align}
\label{eq:QX2matel}
\langle \pi^0(p_\pi)\ell^+(q)|\mathcal{O}_{X,2} |p(p)\rangle =& -\frac{1}{2\sqrt{2}}\,\biggl\{ [v^T(q)\mathcal{C}P_X \Pi(p_\pi)\gamma_\mu P_L]_\gamma G_{\alpha\beta\gamma}(p)[\gamma^\mu]_{\alpha\beta} \nonumber\\
&+ 2[v^T(q)\mathcal{C}P_X]_\delta G_{\alpha \delta\gamma}(p)[\gamma^\mu \Pi(p_\pi)\gamma_\mu P_L]_{\alpha\gamma} \biggr\} \nonumber\\
=&\, \frac{i}{16\sqrt{2}N_c}m_p^2 f_\pi \Bigl[A_2^{XL} M_L + A_2^{XR} M_R \Bigr]\,,
\end{align}
and
\begin{align}
\label{eq:QX1matel}
\langle \pi^0(p_\pi)\ell^+(q)|\mathcal{O}_{X,1} |p(p)\rangle =& \,\frac{1}{3\sqrt{2}}\,\biggl\{ [v^T(q)\mathcal{C}P_X \Pi(p_\pi)\gamma_\mu P_L]_\gamma G_{\alpha\beta\gamma}(p)[\gamma^\mu]_{\alpha\beta} \nonumber\\
&+ 2[v^T(q)\mathcal{C}P_X]_\delta G_{\alpha \delta\gamma}(p)[\gamma^\mu (P_R-2P_L)\Pi(p_\pi)\gamma_\mu P_L]_{\alpha\gamma} \biggr\}\nonumber\\
=& \,\frac{i}{16\sqrt{2}N_c}m_p^2 f_\pi \,\Bigl[A_1^{XL} M_L + A_1^{XR} M_R\Bigr]\,.
\end{align}
Notice the factor $1/\sqrt{2}$ coming from the $\pi^0$ flavour wave-function.
The coefficient functions $A_i^{XY}$ (real, mass dimension 2) depend on $\hat{m}_\pi$, $\hat{m}_\ell$, $\hat{\mu}_\pi$, and the three proton decay constants $f_p$, $\lambda_1$, $\lambda_2$, and are given by
\begin{eqnarray}
    A_1^{XL} &=& \frac{4}{3}\left[ \left(f_p \left(1- 2 \hat{m}_\ell^2 + 2 \hat m_\pi^2\right) 
  +  \lambda_2 \hat \mu_\pi\right) \delta_{XL}
    + 2 f_p \hat m_\ell \delta_{XR} \right],
    \nonumber \\
    A_1^{XR} &=& \frac{4}{3}\left[
    f_p \hat m_\ell~ \delta_{XL} + \left(2 f_p \left(1-2 \hat{m}_\ell^2 + 2 \hat m_\pi^2
    \right) - 3 \lambda_1 \hat \mu_\pi\right) \delta_{XR}
    \right], \nonumber\\[0.05cm]
    A_2^{XL} &=& \left(f_p \left(1-2\hat m_\ell^2 + 2\hat m_\pi^2\right)
    - 3 \lambda_1 -  2\lambda_2 \hat \mu_\pi\right)\, \delta_{XL} 
    + 2 f_p \hat m_\ell\, \delta_{XR}\,,
    \nonumber \\[0.1cm]
    A_2^{XR} &=& \left(2 f_p \left(1-2 \hat m_\ell^2 + 2 \hat m_\pi^2
    \right) + 6 \lambda_1 \hat \mu_\pi\right)\,\delta_{XR} + 
    \left(f_p \hat m_\ell + 3 \lambda_1 \hat m_\ell\right)\,\delta_{XL}\,.
\end{eqnarray}
We further defined
\begin{equation}
    M_X \equiv [v^T(q)\mathcal{C}P_X u_p(p)]\,.
\end{equation}
The sum over the proton and lepton spins yields
\begin{align}
    \frac{1}{2}\sum_{\rm spins} M^\dagger_L M_L = \frac{1}{2}\sum_{\rm spins} M^\dagger_R M_R = m_p^2 \hat{E}_\ell\,,\nonumber\\
    \frac{1}{2}\sum_{\rm spins} M^\dagger_L M_R = \frac{1}{2}\sum_{\rm spins} M^\dagger_R M_L = m_p^2 \hat{m}_\ell\,,
\end{align}
where $\hat{E}_\ell = (1-\hat{m}_\pi^2+\hat{m}_\ell^2)/2$ is the lepton energy normalized by the proton mass.

\begin{figure}[t]
    \centering
    \includegraphics[width=0.45\textwidth]{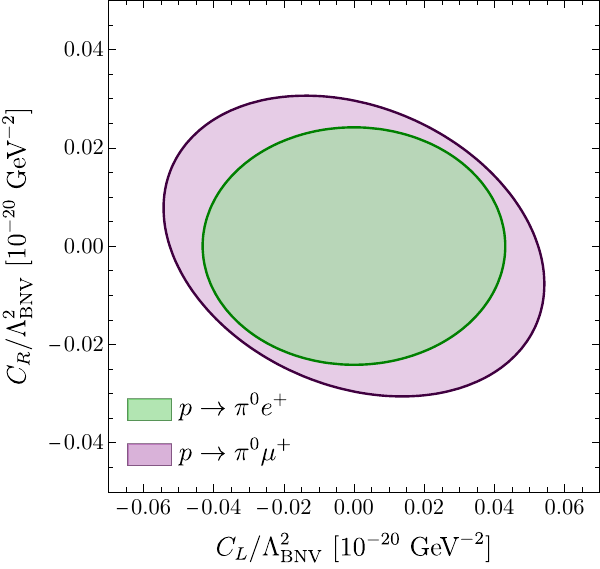}
    \caption{\small Allowed regions for the Wilson coefficients $C_L$ and $C_R$ setting the reference value $\LambdaBNV=10^{10}$~GeV. In green the $p \to \pi^0 e^+$ constraint and in purple the $p \to \pi^0 \mu^+$ constraint.}
    \label{fig:CLCR}
\end{figure}

From~\eqref{eq:Gammappi0ell} we then find
\begin{eqnarray}
    \Gamma(p \to \pi^0 \ell^+) &=& \frac{G_F^2 |V_{ub}|^2|V_{ud}|^2 f_\pi^2 m_p^5}{9216\pi m_b^2 \LambdaBNV^4}\sqrt{(1-\hat{m}_\pi^2)^2 -2\hat{m}_\ell^2 (1+\hat{m}_\pi^2)+\hat{m}_\ell^4}\sum_{X,Y=L,R} C_X C_Y^* \nonumber\\
    &&\hspace*{-2cm}\times \sum_{i,j=1,2} C_i C_j \Bigl[\Bigl(A_i^{XL}A_j^{YL}+A_i^{XR}A_j^{YR}\Bigr)\hat{E}_\ell + \Bigl(A_i^{XR}A_j^{YL}+A_i^{XL}A_j^{YR}\Bigr)\hat{m}_\ell \Bigr]\,.
\label{eq:Gammaptopi0ell}
\end{eqnarray}
Numerically, we obtain
\begin{align}
\label{eq:CLCRbounds}
    \LambdaBNV\Bigl|_{p \to \pi^0 e^+} &> 6.44\cdot 10^{10}~\text{GeV}\,\Bigl(|C_R^e|^2 + 0.0014 \text{Re}[{C_L^e}^* C_R^e] + 0.314 |C_L^e|^2\Bigr)^{1/4}\,,\nonumber\\
    \LambdaBNV\Bigl|_{p \to \pi^0 \mu^+} &> 5.82\cdot 10^{10}~\text{GeV}\,\Bigl(|C_R^\mu|^2 + 0.285 \text{Re}[{C_L^\mu}^* C_R^\mu] + 0.318 |C_L^\mu|^2\Bigr)^{1/4}\,,
\end{align}
where we made explicit the lepton-flavour dependence of the Wilson coefficients. We show in Figure~\ref{fig:CLCR} the allowed region in the $C_L$--$C_R$ plane, in units of $(10^{10}~\text{GeV})^{-2}$, assuming real  Wilson coefficients for the simplicity of presentation.
These bounds are stronger by an order of magnitude than the ones from the processes $p \to \ell^+ \nu_\ell \bar{\nu}$ and $p \to \pi^+ \bar{\nu}$, therefore $C_\nu$ is the largest BNV Wilson coefficient (for the light leptons, $e$ and $\mu$) allowed by proton lifetime limits. 

In Table~\ref{tab:boundssummary} we summarize the lower limits on the scale of BNV obtained in this section for the various right-handed $b$-quark BNV operators.

\begin{table}
    \centering
    \begin{tabular}{c|c|c}
        Exp. constraint & $C$ & $\LambdaBNV/\sqrt{|C|}$ [$10^9$ GeV]\\
        \hline
        {\small dim-8 matching} & $C_Y$ & $>\mathcal{O}(1)$\\
        $p\to e^+ \nu_e \bar{\nu}$ & $C_\nu$ &   $> 6.59$ \\
        $p \to \mu^+ \nu_\mu \bar{\nu}$ & $C_\nu$ & $> 6.86$\\
        $p \to \pi^+ \bar{\nu}$ & $C_\nu$ & $> 3.34$\\
        $p \to \pi^0 e^+$ & $C_R^e$ & $> 64.4$\\
        $p \to \pi^0 e^+$ & $C_L^e$ & $> 48.2$\\
        $p \to \pi^0 \mu^+$ & $C_R^\mu$ & $> 58.2$\\
        $p \to \pi^0 \mu^+$ & $C_L^\mu$ & $> 43.7$\\
        $p \to \ell^+ \nu_\ell \bar{\nu}_\tau$ & $C_{RX}^{3113}$ & $>(0.4\div 1.8)\cdot 10^{-3}$\\
        \hline
    \end{tabular}
    \caption{\small Summary of limits on the scale of right-handed BNV operators containing a $b$ quark, assuming single operator dominance, with $Y=L,R,\nu$ and $X=L,R$.}
    \label{tab:boundssummary}
\end{table}

\section{$B$-meson decay estimates}

Direct searches for baryon-number violating decays have been performed by the BaBar \cite{BaBar:2011yks} and LHCb \cite{LHCb:2022wro} collaborations in exclusive two-body $B$-meson decays to a baryon and a lepton. Upper bounds of about $10^{-9}$ have been placed on the studied branching ratios. In principle, further experimental studies could improve the bounds by a few orders of magnitude.

Inclusive decays of $B$-mesons or $\Lambda_b$-baryons could also be studied experimentally, for example, by tagging a lepton and a single baryon in the final state. For such final states, there is a kinematical window of lepton energy $E_\ell$ that can only be reached by baryon-number violating decays
\begin{equation}
  \frac{m_B}{2} \left(1+\frac{m_\ell^2-4m_N^2}{m_B^2} \right) \le E_\ell \le \frac{m_B}{2} \left(1+\frac{m_\ell^2-m_N^2}{m_B^2} \right)\,,
\end{equation}
where $m_N$ is the mass of the lightest baryon in the final state and $m_B=5.279~\text{GeV}$ is the $B$-meson mass.
The width of the window can be estimated through
\begin{equation}
    \Delta E_\ell\bigl|_{\rm BNV} = \frac{3m_N^2}{2m_B} \approx 250~\text{MeV}\,.
\end{equation}

In order to perform a rough estimate of the branching ratio of BNV $B$-meson decays we consider the inclusive process $\bar{B} \to X \ell$, with a total rate given by
\begin{equation}
\Gamma(\bar{B}\to X\ell) = \frac{1}{2m_B}\int d\Pi_{\rm LIPS} \,|\langle X \ell| \mathcal{H}_{\rm BNV} |\bar{B} \rangle|^2 \approx \frac{4\pi}{2m_B(2\pi)^3}\int_0^{E_\ell^{\rm max}} \!\!dE_\ell\, \frac{E_\ell}{2} \frac{[L \cdot W]}{\LambdaBNV^4}\,,
\end{equation}
and $E_\ell^{\rm max} \approx m_B/2$.
From dimensional analysis, we estimate $L \sim E_\ell$ and $W \sim \pi m_B^3/(16\pi^2)$ for the leptonic and hadronic tensors, respectively.
In the latter we included the loop factor $1/(16\pi^2)$ and a factor of $\pi$ from taking the imaginary part when employing the optical theorem.

Therefore, in the absence of any further suppression by small couplings ($C^j_{\rm BNV}\sim 1$), and by choosing a benchmark conservative bound $\LambdaBNV> 6\cdot 10^{9}$~GeV suggested by~\eqref{eq:leptbounds}, BNV decays of $B$ mesons could have branching fractions up to
\begin{equation}
\label{eq:BRest}
\mathcal{B}(\bar{B}\to X\ell) = \frac{\Gamma(\bar{B}\to X\ell)}{\Gamma_{\rm tot}^B} \approx \frac{m_b^5}{2^{10} \,3\pi^3\LambdaBNV^4 \Gamma_{\rm tot}^B} \approx (8 |V_{cb}| G_F \LambdaBNV^2)^{-2}\lesssim \mathcal{O}(5\cdot 10^{-29})\,,
\end{equation}
where $\Gamma_{\rm tot}^B$ is the total decay width of the $B$-meson. We confirmed this simple estimate by a full computation of the inclusive semi-leptonic branching fraction using the heavy-quark expansion. The  bound~\eqref{eq:BRest} is far from any imaginable direct measurement in the next decades.
Similar (stronger) bounds would apply for $B$ decays involving a strange (charm) quark, 
following from the discussion of Section~\ref{sec:2ndfamily}.

Eq.~\eqref {eq:tauoperator} and the last entry in the table show that the scale of third-generation BNV involving $b$-quarks and $\tau$-leptons can be significantly lower than the one for muons and electrons, which allows 
much larger $B$-meson branching fractions up to
\begin{equation}
\mathcal{B}(\bar{B}\to X \tau)\lesssim \mathcal{O}(10^{-13}\div10^{-15})\,.
\end{equation}
Although this is in the ball-park of the smallest branching fractions for the decay of any particle ever measured, the smaller efficiency of detecting $\tau$-leptons still renders the observation of such decay modes unrealistic in the foreseeable future. 

\section{Conclusions}

We computed the decay rates for the processes $p \to \ell^+ \nu_\ell \bar{\nu}$, $p \to \pi^+ \bar{\nu}$ and $p\to \pi^0 \ell^+$ under the assumption that baryon-number violation is flavour-specific and may occur at a lower scale than the GUT scale, but only if a third-generation quark is involved. Proton decay therefore occurs by a combination of the BNV SMEFT operators and a standard flavour-changing weak interaction, which produce and destroy a virtual $b$-quark in the proton in the most efficient way.
By comparing with the existing experimental bounds for such modes~\cite{Super-Kamiokande:2013rwg,Super-Kamiokande:2014pqx,Super-Kamiokande:2020wjk}, 
we were able to derive bounds of order $10^9-10^{10}$ GeV 
($10^6$ GeV if a $\tau$-lepton is also involved) for the smallest 
energy scale at which third-generation BNV can be present.

The motivation of this work has been to explore whether the scale $\LambdaBNV$
of baryon-number violation could be much below the scale 
$10^{15-16}\,$GeV, which follows from proton stability through BNV operators containing only light quarks, if BNV occurs only when a  third-generation quark is involved. This would offer the possibility that the BNV scale is 
related to the scale $\Lambda_{\rm fl}$ of new flavour physics and 
potentially allow the direct observation of baryon-number violating $b$-hadron decays, which display unique signatures. This is particularly appealing if $\Lambda_{\rm fl}$ is also the scale, at 
which the special role of the third family is imprinted on the 
SM flavour sector. Our calculations and estimates show, however, that 
while $\LambdaBNV$ can be six orders (or even nine orders in the case 
of $\tau$-lepton final states) below $10^{15-16}\,$GeV, 
$\LambdaBNV\sim \Lambda_{\rm fl}\sim$ TeV is ruled out, since proton 
stability constrains the third-generation BNV operators indirectly. 
While $10^{15}\,\mbox{GeV}\gg\LambdaBNV\gg\Lambda_{\rm fl}\sim$ TeV 
is technically an option, the implicit assumption that the 
third generation is singled out at a scale far above 
$\Lambda_{\rm fl}$ does not appear to be very attractive from the point of model-building. 
In any case, the obtained lower limits on the scale of highly generation-dependent BNV favouring the third family exclude the possibility of observing such BNV directly in $b$-hadron decays in any presently conceivable experiment.

\subsubsection*{Acknowledgements}
We thank Vladimir Braun for discussions. 
MB thanks the Galileo Galilei Institute for Theoretical Physics for hospitality and the INFN for
support during the completion of this work. 
A.A.P. would like to thank the Excellence Cluster ORIGINS
(funded by the Deutsche Forschungsgemeinschaft  under Germany's Excellence Strategy ``EXC 2094'' 390783311) 
for hospitality.
GF thanks the Department of Physics and Astronomy of the University of South Carolina for hospitality during the initial phases of this project.
This research was supported in part by the Deutsche Forschungsgemeinschaft (DFG, German Research Foundation) through the Sino-German Collaborative Research Center TRR110 ``Symmetries
and the Emergence of Structure in QCD'' (DFG Project-ID 196253076, NSFC Grant No. 12070131001, - TRR 110).
A.A.P. was also supported in part by the DOE grant DE-SC0024357.

\appendix

\section{Constraints on light-flavoured BNV operators}
\label{sec:lightBNV}

As a reference value for the scale of the Wilson coefficients of BNV operators with only first family quarks  (``light-BNV'' operators), 
\begin{equation}
   Q^{111}_{XY} = \varepsilon^{abc} [\widetilde{d}^a P_X u^b]\,[\widetilde{u}^c P_Y \ell]\,,
\end{equation}
we consider the constraint from $p \to \pi^0 \ell^+$ measurements~\cite{Super-Kamiokande:2020wjk}. The branching ratio can be easily computed employing (\ref{eq:simprel}) and~\cite{Haisch:2021hvj} 
\begin{equation}
    \langle \pi^0 \ell^+(q)| Q_{XY}^{111}|p(p)\rangle = i \,[v^T(q) \mathcal{C}P_Y \Bigl(W^0_{XY} + \frac{\slashed{q}}{m_p}W^1_{XY} \Bigr)u_p(p)]\,,
\end{equation}
where $u_p$ ($v$) is the proton (anti-lepton) spinor. 
From light-cone sum-rules~\cite{Haisch:2021hvj} 
one obtains the numerical values
\begin{align}
    W^0_{LL} = W^0_{RR} &= +0.084 \pm 0.021~\text{GeV}^2\,, \qquad W^1_{LL} = W^1_{RR} = -0.068\pm 0.023~\text{GeV}^2\,,\nonumber\\
    W^0_{LR} = W^0_{RL} &= -0.118 \pm 0.030~\text{GeV}^2\,, \qquad W^1_{LR} = W^1_{RL} = +0.14\pm 0.06~\text{GeV}^2
\end{align}
for the form factors. Assuming single operator dominance, the decay rate reads
\begin{equation}
\label{eq:Gammalight}
    \Gamma(p \to \pi^0 \ell^+) = \frac{|C_{XY}^{111}|^2}{32\pi \LambdaBNV^4} m_p (W^0_{XY})^2 + \mathcal{O}\left(\frac{m_\ell^2}{m_p^2}, \frac{m_\pi^2}{m_p^2} \right) \,.
\end{equation}
The measurements summarized in Table~\ref{tab:input} of Appendix~\ref{sec:inputs} translate into the bounds (restoring the lepton generation index)
\begin{align}
\label{eq:lightbound}
    \frac{\LambdaBNV}{\sqrt{|C_{LL}^{1111}|}} &= \frac{\LambdaBNV}{\sqrt{|C_{RR}^{1111}|}} > 3.0\cdot 10^{15}~\text{GeV}\,, \qquad   \frac{\LambdaBNV}{\sqrt{|C_{LR}^{1111}|}}=\frac{\LambdaBNV}{\sqrt{|C_{RL}^{1111}|}} > 3.5\cdot 10^{15}~\text{GeV}\,,\nonumber\\
    \frac{\LambdaBNV}{\sqrt{|C_{LL}^{1112}|}} &= \frac{\LambdaBNV}{\sqrt{|C_{RR}^{1112}|}} > 2.7\cdot 10^{15}~\text{GeV}\,, \qquad   \frac{\LambdaBNV}{\sqrt{|C_{LR}^{1112}|}}=\frac{\LambdaBNV}{\sqrt{|C_{RL}^{1112}|}} > 3.2\cdot 10^{15}~\text{GeV}\,,
\end{align}
which indeed implies $\LambdaBNV \approx \Lambda_{\rm GUT}$ for the BNV scale, assuming $\mathcal{O}(1)$ Wilson coefficients.
The numerical differences between the electron and muon channel are a consequence of the different experimental constraints.
These bounds motivate the assumption that the Wilson coefficients of operators involving only first-family quarks are negligible at the electroweak scale in a scenario where $\LambdaBNV\ll \Lambda_{\rm GUT}$.

The two operators $Q_{L\nu}$ and $Q_{R\nu}$ involving the neutrino mediate the decay $p\to \pi^+ \bar{\nu}$ at tree-level, which is less constrained experimentally.
However, assuming the hadronic form factor for $p \to \pi^+$ to be of the same order as for $p \to \pi^0$, the bounds~\eqref{eq:lightbound} only have to be rescaled by the factor
\begin{equation}
   \biggl[\frac{\Gamma(p \to \pi^0 e^+)}{\Gamma(p \to \pi^+ \bar{\nu})}\biggl|_{\rm exp}\biggr]^{1/4} = 0.36\,,
\end{equation}
which allows us to treat the two operators containing the neutrino on the same level as the other four.
Similar bounds would apply to $Q^{211}_{XY}$ since the decays $p\to K^0 \ell^+$ and $p \to K^+ \bar{\nu}$ are also severely constrained by data~\cite{Super-Kamiokande:2022egr,Super-Kamiokande:2005lev}. The above estimates are in agreement with the comprehensive analysis of bounds for light-flavoured operators recently presented in~\cite{Beneito:2023xbk}.

\section{Numerical inputs}
\label{sec:inputs}

\begin{table}[ht]
\centering
\begin{tabular}{|c|c|}
\hline
\multicolumn{2}{|c|}{}\\[-4.0mm]
\multicolumn{2}{|c|}{Masses~\cite{PDG}}\\[1mm]
\hline
&\\[-4.0mm]
$m_b = 4.8~\text{GeV}$ & $m_p=0.938~\text{GeV}$ \\[1mm]
$m_e = 0.511~\text{MeV}$ & $m_\mu = 105.66~\text{MeV}$\\[1mm]
$m_\tau = 1.777~\text{GeV}$ & $m_\pi = 139.57~\text{MeV}$\\[1mm]
$m_u(2~\text{GeV}) = 2.16~\text{MeV}$ & $m_d(2~\text{GeV}) = 4.67~\text{MeV}$\\[1mm]
\hline
\multicolumn{2}{|c|}{}\\[-4.0mm]
\multicolumn{2}{|c|}{Coupling constants~\cite{PDG}}\\[1mm]
\hline
&\\[-4.0mm]
$G_F = 1.1663788 \cdot 10^{-5}~\text{GeV}^{-2}$ &  $\as^{(5)}(m_Z) = 0.1179$ \\[1mm]
\hline
\multicolumn{2}{|c|}{}\\[-4.0mm]
\multicolumn{2}{|c|}{CKM matrix elements}\\[1mm]
\hline
&\\[-4.0mm]
$|V_{ud}| = 0.9737$~\cite{ParticleDataGroup:2020ssz}  & $|V_{ub}| =3.77\cdot 10^{-3}$~\cite{Leljak:2021vte} \\[1mm]
\hline
\multicolumn{2}{|c|}{}\\[-4.0mm]
\multicolumn{2}{|c|}{Decay constants (at 2 GeV for the proton~\cite{RQCD:2019hps})}\\[1mm]
\hline
&\\[-4.0mm]
$f_p = 3.54^{+0.06}_{-0.04}\cdot 10^{-3}~\text{GeV}^2$ & $\lambda_1 = -(44.9^{+4.2}_{-4.1})\cdot 10^{-3}~\text{GeV}^2$\\[1mm]
$\lambda_2 = 93.4^{+4.8}_{-4.8}\cdot 10^{-3}~\text{GeV}^2$ & $f_\pi = 130.2~\text{MeV}$~\cite{FlavourLatticeAveragingGroupFLAG:2021npn} \\[1mm]
\hline
\multicolumn{2}{|c|}{}\\[-4.0mm]
\multicolumn{2}{|c|}{Wilson coefficients~\cite{Gorbahn:2004my}}\\[1mm]
\hline
&\\[-4.0mm]
$C_1(1~\text{GeV}) = -0.829$ &  $C_2(1~\text{GeV}) = 1.050$ \\[1mm]
\hline
\multicolumn{2}{|c|}{}\\[-4.0mm]
\multicolumn{2}{|c|}{Partial lifetimes (90\% CL) \& decay rates}\\[1mm]
\hline
&\\[-4.0mm]
$\tau_{p \to e^+ \nu \nu} > 1.7\cdot 10^{32}~\text{yr}$~\cite{Super-Kamiokande:2014pqx} & $\tau_{p \to \mu^+ \nu \nu} > 2.2\cdot 10^{32}~\text{yr}$~\cite{Super-Kamiokande:2014pqx}\\[1mm]
$\tau_{p \to \pi^+ \bar{\nu}} > 3.9\cdot 10^{32}~\text{yr}$~\cite{Super-Kamiokande:2013rwg} & $\tau_{p \to \pi^0 \mu^+} > 1.6\cdot 10^{34}~\text{yr}$~\cite{Super-Kamiokande:2020wjk} \\[1mm]
$\tau_{p \to \pi^0 e^+} > 2.4\cdot 10^{34}~\text{yr}$~\cite{Super-Kamiokande:2020wjk} & $\Gamma_{\rm tot}^B = 4.4\cdot 10^{-13}~\text{GeV}$~\cite{PDG}\\[1mm]
\hline
\end{tabular}
\caption{Numerical inputs used for the determination of the bounds on $\LambdaBNV$ and $\mathcal{B}^B_{\rm BNV}$.}
\label{tab:input}
\end{table}

\noindent We list in Table~\ref{tab:input} the inputs used for our numerical analysis. We do not quote most of the uncertainties since our results are in the form of bounds (coming from $90\%$ CL bounds on proton lifetimes).
The partial lifetimes are defined as
\begin{equation}
    \tau_{p \to f} = \frac{1}{\Gamma(p \to f)}\,,
\end{equation}
with the unit conversion factor $1~\text{yr} = 4.79434 \cdot 10^{31}~\text{GeV}^{-1}$.
For the strong coupling  $\as(\mu)$ in the $\overline{\rm MS}$ scheme we use \texttt{RunDec}~\cite{Herren:2017osy} with three-loop running, and decouple the charm quark at $\mu_c =3~\text{GeV}$ such that $n_f=3$ between 2 and 1~GeV.
The value of\footnote{Note that the expression and value of $\mu_\pi$ applies to both the charged and neutral pion, see Section 3.7.2 of~\cite{Beneke:2000ry}.}
\begin{equation}
\label{eq:mupi}
    \mu_\pi = \frac{m_\pi^2}{m_{ud}(1~\text{GeV})} = 2.315~\text{GeV}\,,
\end{equation}
is computed at 1 GeV by evolving the quark masses $m_{ud}\equiv m_u+m_d$ with
\begin{equation}
m_{ud}(\mu) = \biggl(\frac{\as(\mu)}{\as(\mu_0)} \biggr)^{4/\beta_0}m_{ud}(\mu_0)\,, 
\end{equation}
and 
\begin{equation}
    \beta_0 = 11 - \frac{2}{3}n_f = 9\,.
\end{equation}

For the proton decay constants we use the values at 2 GeV from~\cite{RQCD:2019hps}, and evolve them to 1 GeV with the relations~\cite{Anikin:2013aka,Krankl:2011gch}
\begin{equation}
f_p(\mu) = \biggl(\frac{\as(\mu)}{\as(\mu_0)} \biggr)^{\!\frac{2}{3\beta_0}}f_p(\mu_0)\,,\qquad \lambda_i(\mu) = \biggl(\frac{\as(\mu)}{\as(\mu_0)} \biggr)^{\!-\frac{2}{\beta_0}}\lambda_i(\mu_0)\,,
\end{equation}
for $i=1,2$, obtaining the values $f_p(1~\text{GeV}) = 3.67\cdot 10^{-3}~\text{GeV}^2$, $\lambda_1(1~\text{GeV}) = -40.5\cdot 10^{-3}~\text{GeV}^2$ and $\lambda_2(1~\text{GeV}) = 84.2\cdot 10^{-3}~\text{GeV}^2$.


\section{Pion and proton decay matrix elements}
\label{sec:appproj}

\subsection{Pseudoscalar meson}
\label{sec:pseudomes}
For the pseudoscalar meson we consider the positively charged pion. We wish to parametrize the matrix element of the local field product $\bar{u}_\alpha d_\beta$ between the vacuum and the $\pi^+$ state in terms of the most general Dirac structures:
\begin{equation}
M_{\beta\alpha}(p)\equiv \langle \pi^+(p)|\bar{u}_\alpha d_\beta | 0\rangle = S \mathbb{1}_{\beta \alpha} + V\slashed{p}_{\beta\alpha} + P\gamma^5_{\beta \alpha} + A[\slashed{p}\,\gamma^5]_{\beta\alpha}\,,
\end{equation}
where the fields are evaluated at $x=0$ and the antisymmetric tensor term vanishes since there is only one Lorentz vector to contract the indices.

We can apply a parity transformation in order to further constrain the basis by writing 
\begin{align}
\label{eq:Pcond}
M_{\beta\alpha}(p) &= \langle \pi^+(p) |P^{-1}P\bar{u}_\alpha P^{-1}P d_\beta P^{-1} P |0\rangle\nonumber\\
 &= -\langle \pi^+(\tilde{p})| [\bar{u}\gamma^0]_\alpha [\gamma^0 d]_\beta |0\rangle = - [\gamma^0 M(\tilde{p})\gamma^0]_{\beta\alpha}\,,
\end{align}
where we used the fact that the pion has negative parity and $\tilde{p}^\mu = (p^0,-\vec{p}\,)$. Requiring that the condition~\eqref{eq:Pcond} is satisfied, we find $S=0$ and $V=0$. The definition of the pion decay constant, 
\begin{equation}
\label{eq:decconst}
\langle \pi^+(p) |\bar{u}(x) \gamma^\mu \gamma^5 d(x)|0\rangle = -i f_\pi p^\mu e^{i p \cdot x}\,, 
\end{equation}
implies $A = \frac{i}{4}f_\pi$.
Taking the derivative of~\eqref{eq:decconst} on both sides, and by applying the equation-of-motion of the fields we get
\begin{equation}
\langle \pi^+(p) |\bar{u}(x) (\overleftarrow{\slashed{\partial}} \gamma^5 - \gamma^5 \slashed{\partial}) d(x)|0\rangle = i (m_d+m_u)\langle \pi^+(p) |\bar{u}(x) \gamma^5 d(x)|0\rangle = f_\pi m_\pi^2 e^{i p \cdot x}\,,
\end{equation}
therefore
\begin{equation}
\langle \pi^+(p) |\bar{u}(x) \gamma^5 d(x)|0\rangle = -i f_\pi \mu_\pi e^{i p\cdot x}\,,
\end{equation}
where $\mu_\pi = m_\pi^2/(m_d+m_u)$. This relation implies $P = -\frac{i}{4}f_\pi \mu_\pi$. For uncontracted colour indices the matrix element is proportional to $\delta^{ab}$, since the meson is a colour singlet. 
Hence the final result for the matrix element is 
\begin{equation}
\label{eq:projpi}
\langle  \pi^+(p)|\bar{u}_\alpha^a(0) d_\beta^b(0) | 0\rangle = \frac{i\delta^{ab}}{4N_c}f_\pi (\slashed{p}\gamma^5 - \mu_\pi \gamma^5)_{\beta\alpha}\,.
\end{equation}

\subsection{Proton}
\label{sec:protdeco}
The matrix element of interest can be parametrized as
\begin{equation}
\label{eq:protdecomp}
G_{\alpha\beta\gamma}(p) \equiv \langle 0| \varepsilon^{abc} \widetilde{u}^a_\alpha\, u^b_\beta \, d^c_\gamma \, |p(p)\rangle = \sum_{i,j} f^{ij} M^i_{\beta \alpha}(p) [\Gamma^j u_p(p)]_\gamma\,,
\end{equation}
where $u_p(p)$ is the proton spinor and $i,j$ label all the possible independent Dirac structures, as was done in Section~\ref{sec:pseudomes}:
\begin{align}
\label{eq:GammaiMi}
    \Gamma^j &= \{\mathbb{1}, \gamma^5, \gamma_\mu,i\gamma_\mu \gamma^5, \sigma_{\mu\nu}\}\,,\nonumber\\
    M^i(p) &= \{\mathbb{1}, \gamma^5, \slashed{p}, \slashed{p}\gamma^5, \gamma^\mu, \gamma^\mu \gamma^5,\sigma^{\mu\nu}, p_\nu \sigma^{\mu\nu},\varepsilon^{\mu\nu\rho\sigma}\sigma_{\rho\sigma}\}\,,
\end{align}
with the constraint that the contraction between $M^i$ and $\Gamma^j$ is a Lorentz scalar.
Notice that $\Gamma^j$ is independent of $p$ since we can use the equation-of-motion of the spinor $u_p(p)$. Each term is multiplied by a non-perturbative parameter $f^{ij}$.
In~\eqref{eq:protdecomp} one would also expect further terms with  indices $\alpha,\beta,\gamma$ permuted on the right-hand side; however, they can be reabsorbed in the already present term by using the closure relation / Fierz transformation
\begin{equation}
    \delta_{\alpha\beta}\delta_{\gamma\delta} = \frac{1}{4}\sum_{j\neq \sigma_{\mu\nu}} \Gamma^j_{\alpha \delta}\Gamma^j_{\gamma\beta} + \frac{1}{8}[\sigma^{\mu\nu}]_{\alpha \delta}[\sigma_{\mu\nu}]_{\gamma\beta}\,,
\end{equation}
where the Lorentz indices are contracted between the two $\Gamma^j$.

We can use parity to constrain the building blocks, by inserting $PP^{-1}$ between the fields in the matrix element defining $G_{\alpha\beta\gamma}$.
The resulting relation is
\begin{equation}
\label{eq:Pconst}
M^i_{\beta \alpha}(p) [\Gamma^j u(p)]_\gamma = -[\gamma^0 M^i(\tilde{p}) \gamma^0]_{\beta\alpha}[\gamma^0 \Gamma^j \gamma^0 u_p(p)]_\gamma\,,
\end{equation}
where we used $u_p(\tilde{p}) = \gamma^0 u_p(p)$. The relative minus sign comes from the parity transformation on the $\widetilde{u}$ field. 
Furthermore, by anticommuting the two up-quark fields in the definition of $G_{\alpha\beta\gamma}$ we get the relation
\begin{equation}
G_{\alpha\beta\gamma} =\mathcal{C}_{\alpha\rho} G_{\sigma\rho\gamma} \mathcal{C}_{\sigma\beta}\,,
\end{equation}
where $\mathcal{C}$ is the charge conjugation matrix with properties~\eqref{eq:Cprop}.
Using the fact that the Dirac structures labelled by $j$ are independent, we arrive at 
\begin{equation}
\label{eq:Cconst}
M^i(p) = \mathcal{C} M^i(p)^T \mathcal{C}\,,
\end{equation}
which forces $M^i$ in~\eqref{eq:GammaiMi} to be either the vector or the tensor Dirac structures.

Next we exploit the conditions~\eqref{eq:Pconst} and~\eqref{eq:Cconst} to find all the allowed Dirac structures contributing to $G_{\alpha\beta\gamma}(p)$.
We start by listing the elements of $M^i_{\beta\alpha}(p)$ satisfying the constraint~\eqref{eq:Cconst}:
\begin{equation}
\label{eq:MiCconst}
 \slashed{p}, \gamma^\mu, p_\nu \sigma^{\mu\nu}, \sigma^{\mu\nu}, \varepsilon^{\mu\nu\rho\sigma}\sigma_{\mu\nu}\,.
\end{equation}
For each of these structures we identify the possible $\Gamma^j$ which fulfil the constraint~\eqref{eq:Pconst} and form a Lorentz scalar with $M^i(p)$. 
We find that for each allowed $M^i(p)$ there is only one allowed $\Gamma^j$, resulting in 
\begin{equation}
    M^i(p)\otimes \Gamma^j = \{ \slashed{p} \otimes \gamma^5, \gamma^\mu \otimes \gamma_\mu \gamma^5, p_\nu \sigma^{\mu\nu}\otimes \gamma_\mu \gamma^5, \varepsilon^{\mu\nu\rho\sigma}\sigma_{\mu\nu}\otimes \sigma_{\rho\sigma}\}\,,
\end{equation}
where we used $\slashed{\tilde{p}}=\slashed{p}^\dagger$ and the short-hand $M^i(p)\otimes \Gamma^j = M^i_{\beta\alpha}(p) \Gamma^j_{\gamma\delta}$. 
Therefore the most general decomposition for the proton projector is
\begin{equation}
G_{\alpha\beta\gamma}(p) = V_P \slashed{p}_{\beta\alpha} [\gamma^5 u(p)]_\gamma + \Bigl(V_A \gamma_\rho + T_A p^\sigma \sigma_{\rho\sigma} \Bigr)_{\beta\alpha} [\gamma^\rho \gamma^5 u(p)]_\gamma + T_T [\sigma_{\rho\sigma}]_{\beta\alpha} [\sigma^{\rho \sigma}\gamma^5 u(p)]_\gamma\,,
\end{equation}
where we renamed the $f^{ij}$ and used
\begin{equation}
\frac{i}{2}\varepsilon^{\mu\nu\rho\sigma}\sigma_{\rho\sigma}=\sigma^{\mu\nu} \gamma^5\,.
\end{equation}
The structures are in agreement with Eq. (3.11) of~\cite{Braun:2000kw}.

The parameters $V_P$, $V_A$, $T_A$ and $T_T$ are related to 
$f_p,f_p^T,\lambda_1,\lambda_2$ computed by the RQCD lattice collaboration~\cite{RQCD:2019hps} (see numerical values in Table~\ref{tab:input}) by  
\begin{align}
V_P &= -\frac{f_p}{4}\,,\qquad V_A = \frac{m_p}{16}(\lambda_1 - f_p)\,,\nonumber\\
T_A &= -\frac{i}{4}f_p^T\,,\qquad T_T = \frac{m_p}{96}(\lambda_2 - 6f_p^T)\,.  
\end{align}
Isospin symmetry implies $f_p^T = f_p$. In terms of these parameters, 
the full proton matrix projector is given by
\begin{align}
G_{\alpha\beta\gamma}(p) =& -\frac{f_p}{4}\biggl(\slashed{p}_{\beta\alpha}[\gamma^5 u_p(p)]_\gamma + i p^\nu [\sigma_{\rho\nu}]_{\beta\alpha} [\gamma^\rho \gamma^5 u_p(p)]_\gamma  \biggr) \nonumber\\
&+ \frac{m_p}{16}(\lambda_1-f_p)[\gamma_\rho]_{\beta\alpha}[\gamma^\rho \gamma^5 u_p(p)]_\gamma +  \frac{m_p}{96}(\lambda_2 -6f_p)[\sigma_{\rho\sigma}]_{\beta\alpha} [\sigma^{\rho \sigma}\gamma^5 u_p(p)]_\gamma \,.
\end{align}

\bibliographystyle{JHEP}
\bibliography{refs.bib}
\end{document}